\documentclass[11pt,a4paper]{article}
\usepackage{jheppub}
\usepackage[utf8]{inputenc}
\usepackage[english]{babel}
\usepackage{amsmath}
\usepackage{amsfonts}
\usepackage{amssymb}

\newcommand{\be}{\begin{equation}}
\newcommand{\ee}{\end{equation}}
\newcommand{\bs}{\begin{equation}\begin{split}}
\newcommand{\es}{\end{equation}\end{split}}

\newcommand{\ti}[1]{{\text{\tiny{\textit{#1}}}}}
\newcommand{\spin}[1]{\sigma_ {#1}}
\newcommand{\pat}[2]{\xi ^{#1}_{#2}}

\newtheorem{Remark}{Remark}
\newtheorem{Theorem}{Theorem}
\newtheorem{Proposition}{Proposition}
\newtheorem{Corollary}{Corollary}
\newtheorem{Definition}{Definition}

\title{A relativistic extension of Hopfield neural networks via the mechanical analogy}
\author[a,b,c]{Adriano Barra,}
\author[a,b]{Matteo Beccaria%
}
\author[a,b]{and Alberto Fachechi}
\affiliation[a]{Dipartimento di Matematica e Fisica Ennio De Giorgi,\\
Università del Salento, Lecce, Italy}
\affiliation[b]{INFN, Istituto Nazionale di Fisica Nucleare, Sezione di Lecce, Italy}
\affiliation[c]{GNFM-INdAM, Gruppo Nazionale per la Fisica Matematica, Sezione di Lecce, Italy}
\emailAdd{adriano.barra@unisalento.it}
\emailAdd{matteo.beccaria@le.infn.it}
\emailAdd{alberto.fachechi@le.infn.it}

\abstract{
We propose a modification of the cost function of the Hopfield model whose salient features shine in its Taylor expansion and result in more than pairwise interactions with alternate signs, suggesting a unified framework for handling both with deep learning and network pruning. In our analysis, we heavily rely on the Hamilton-Jacobi correspondence relating the statistical model with a mechanical system. In this picture, our model is nothing but the relativistic extension of the original Hopfield model (whose cost function is a quadratic form in the Mattis magnetization which mimics the non-relativistic Hamiltonian for a free particle). We focus on the low-storage regime and solve the model analytically by taking advantage of the mechanical analogy, thus obtaining a complete characterization of the free energy and the associated self-consistency equations in the thermodynamic limit. On the numerical side, we test the performances of our proposal with MC simulations, showing that the stability of spurious states (limiting the capabilities of the standard Hebbian construction) is sensibly reduced due to presence of unlearning contributions in this extended framework.
}

\begin{document}
\maketitle

\section{Introduction}

In the past few years, thanks both to progresses in hardware (possibly due to the novel generation of GPU computing architectures \cite{GPU1,GPU2}), as well as in software  ({\it e.g.} driven by results from Google and related companies as Deep Mind), we assisted at the rise of a novel and more powerful generation of Artificial Intelligence, whose most popular contributions perhaps lie in the impressive adaptive capability of {\em Deep Learning}  \cite{DL1,DL2} and in the creative compositional episodes during {\em sleeping and dreaming} \cite{Guardian,DeepDream,Monasson}.
\newline
\newline
Regarding the former, deep learning, beyond literally grounding Artificial Intelligence within the theoretical framework of {\em disordered statistical mechanics} \cite{Hopfield,Amit}, Hopfield recently  offered also a connectionist perspective where the high skills of deep learning machines could possibly be framed \cite{HopfieldNew}: the route he proposes is via many-body extensions of his celebrated pairwise model for feature retrieval and pattern recognition \cite{Amit,Coolen,Hopfield}. The core of the idea is as elegant as simple, as already pointed out by Hopfield and Tank in the past \cite{HopfieldTank}: consider $N$ on/off neurons $\sigma_i = \pm 1$ (where $i$ labels neurons from $1$ to $N$), if we want them to retrieve one out of $P$ random patterns $\bold{\xi}^{\mu}$ ($\mu$ labeling the patterns from $1$ to $P$) and we want to describe this property via a cost-function $H(\boldsymbol{\sigma}\vert\boldsymbol{\xi})$ that resembles Hamiltonians in Physics (such that the minima of the Hamiltonian would match the patterns themselves \cite{Amit,MPV,Treves}), the simplest and most natural guess would be summing all  the squared scalar products between the neurons and the patterns, i.e., $H(\boldsymbol{\sigma}\vert\boldsymbol{\xi}) \propto - \sum_{\mu}^P (\boldsymbol{\sigma}\cdot \boldsymbol{\xi}^{\mu})^2$: for large $N$, patterns become orthogonal\footnote{With the term {\em orthogonal} we mean that $\lim_{N \to \infty} N^{-1}\bold{\xi}^{\mu}\bold{\xi}^{\nu} = \lim_{N \to \infty} N^{-1} \sum_i^N \xi_i^{\mu}\xi_i^{\nu} = \delta_{\mu\nu}$. However, at finite $N$, this is a $N$-long sum of terms whose probability of being $\pm1$ is
 one half: it is a random walk with zero mean and variance $N$, hence spurious correlations are expected to vanish $\propto 1/\sqrt{N}$.} and if the state vector $\boldsymbol{\sigma}$ is uncorrelated with all of them, each parenthesized term would be very small, but if the state network $\boldsymbol{\sigma}$ retrieves one of the $P$ patterns (i.e., it becomes highly correlated with the pattern), then its contribution in the summation would be no longer negligible and the pattern would actually act as an attractor for any reasonable stochastic neural dynamics obeying Detailed Balance \cite{Amit,Coolen}.
\newline
The robustness of this argument lies in the usage of just local convexity of the cost function and can be generalized straightforwardly beyond the parabolic approximation coded by the pairwise interactions, for instance by including also the even higher order contributions (the so-called {\em P-spin} terms): remembering that, from the connectionist perspective \cite{Amit,connessionisti1,connessionisti2}, memories lie in the connections (they are stored in the slowly evolving values of the synapses as learning takes place), clearly adding more and more P-spin contributions to the Hamiltonian adds more and more synapses where information can be filed.
\newline
\newline
Regarding the latter, sleeping and dreaming, while the idea that sleep can actually consolidate memories by discarding fake information accidentally stored is more than two centuries old \cite{200}, its inspection has improved significantly since the discovery of REM sleep in the fifties as pioneered by Aserinsky and Kleitman \cite{REM,unlearning4}. Models of REM sleep have already been framed within a neural network perspective by Crick and Mitchinson \cite{Crick} and Hopfield himself \cite{HopfieldUnlearning}: the whole gave rise to the theory of {\em unlearning} in neural networks \cite{unlearning0,unlearning1,unlearning2,unlearning3}, a very nice idea to remove spurious states from the landscape of retrievable patterns: one of the aim of the present work is to show that unlearning in Hebbian models can be used for pruning these networks.
\newline
Up to nowadays, however, these two branches of neural networks -deep learning and unlearning- never merged and the main aim of the present work is to obtain (in a very natural way) a unified model able to handle both of these features at once.
\newline
\newline
With this goal in our mind, we now turn to methodologies rather than subjects. Initially with the focus on spin glasses \cite{GuerraSumRule,Barra-Guerra-HJ}, in the past ten years, several contributions linking the statistical mechanical recipes to analytical mechanical approaches appeared in the Literature (see {\it e.g.} \cite{Agliari-Dantoni,Agliari-Isopi,Tonio1,Tonio2,BarraBipPDE,Barra-quantum} and references therein). To fix the ideas, let us consider the Curie-Weiss model \cite{Barra-JSP0,genovese} as the simplest example of a pairwise spin Hamiltonian (that can be thought of as an Hopfield model with a unique pattern under the Mattis gauge \cite{Amit}): it has been proved that its free energy obeys a standard  ({\it i.e.} {\em classical}) Hamilton-Jacobi PDE in the space of the coupling and external field (where the former plays as time, and the latter as space in this mechanical analogy): to infer statistical properties on the Curie-Weiss network, we can thus use this mechanical analogy and study a fictitious particle of unitary mass classically evolving in this $1+1$ space-time (under the proper PDE derived from the statistical mechanical framework). Its properties, once translated back in the original setting, recover sharply all the results of the standard route \cite{Barra-JSP0,genovese}.  It is thus possible to perform a whole statistical mechanical analysis of the model (e.g., obtain an explicit expression for the free energy, the self consistent equations for the order parameters, etc.) upon relying solely on techniques typical of {\em analytical mechanics} \cite{Arnold}.
\newline
Here, once this framework is adapted to the Hopfield model, we show that such an analogy calls for a very natural generalization of the Hopfield cost-function (or {\em Hamiltonian} \cite{Amit}), that is simply its relativistic extension. Indeed, we will see how, within the {\em mechanical analogy}, moving from the classical kinetic energy to its relativistic counterpart, Hamilton-Jacobi PDE reduces to the energy-momentum relation \cite{Cabibbo} and the free energy -that coincides with the {\em action} in this equivalence- turns out to obey a relativistic Least Action Principle. While the classical expression for the kinetic energy is a second-order monomial (in the Mattis order parameters), {\it i.e.} the Hopfield cost-function, its relativistic expression is not such but can be Taylor-expanded in all the monomials. We remark that these turn out to be exactly solely the even ones and with alternate signs: the relativistic extension naturally suggested by the mechanical analogy accounts for the higher-order contributions (beyond Hopfield's pairwise one), hence of potential interest for researcher in Deep Learning, yet being appealing for researches in unlearning, given the alternation of signs in the series. In those regards  ({\it i.e.} focusing on Machine Learning), it is worth pointing out that we will work always with randomly generated pattern's entries as in standard Amit-Gutfreund-Sompolinsky theory \cite{Amit}: while this choice barely resembles real situations, however, at a more abstract level, by a standard Shannon-Fano compression argument it is immediate to realize that if the network is able to cope with these $P$ entirely random patterns, it will be certainly able to handle (at least) the same amount of {\em structured} patterns (where correlations are expected and thus their compression would eventually save memory for further storage).
\newline
As a final remark, we stress that while numerical and heuristic explorations have been largely exploited in the Computer Science Literature in the past decade, here we aim to reach sharp statements, with rigorous and well controllable results. Indeed, along this perspective, important contributions already appeared in the Mathematical and Theoretical Physics Literature (see {\it e.g.} \cite{BarraHowGlassy,Barra-quantum,Benarous,Bovier1,Bovier2,Bovier3,Bovier4,Tala1,Tala2,Tirozzi1} and references therein). In these regards, in this first paper we force our focus to the low-storage analysis of the model, namely we study properties the model naturally possesses when the number of patterns (or features) to learn and retrieve grows sub-linearly with the number of neurons dedicated to the task: from the mathematical perspective, this regime is much more controllable as the {\em glassiness} hidden in the model becomes negligible \cite{Amit,Coolen}.
\newline
\newline
The paper is structured as follows:
\newline
We spend the next section (Sec.$2$) for a streamlined introduction to machine learning and neural networks, in particular linking {\em learning} and {\em retrieval} in their simplest representation: we use restricted Boltzmann machines (RBM)  as prototypes for learning\footnote{We remark that deep learning structures can be built by properly hierarchically nestled chains of RBMs one into another \cite{Hinton1}.} and generalized Hopfield models (GHM) as paradigms for retrieval.  We revise, by using a standard Bayes argument, how the features learnt during the training stage by a restricted Boltzmann machine play as patterns to be recognized in the future, highlighting how -de facto- pattern recognition happens via a standard Hebbian kernel typical of Hopfield retrieval: this is a recently understood bridge glimpsed in Disordered Systems Community \cite{Agliari-Dantoni,Agliari-PRL1,BarraEquivalenceRBMeAHN,Barra-RBMsPriors,Mezard,Monasson} that still deserves to be outlined (especially because, due to this bridge, unlearning in Hopfield network can be connected to pruning in Boltzmann machines). Finally we briefly comment on spurious states and their intrinsic genesis in the Hopfield framework.
\newline
Then we move on to build our mathematical framework, {\it i.e.} the mechanical analogy for neural networks: as a benchmark, in Section $3$ we show the analogy at the classical level, namely we consider the standard statistical mechanical package related to the original pairwise Hopfield model (whose mechanical counterpart lies in classical kinetics) and we give a complete picture of its properties, re-obtaining all the well-known existing results. In Section $4$, we extend the analogy to include higher order (P-spin) contributions to the Hopfield Hamiltonian (whose mechanical counterpart lies in special relativity) and we obtain an exhaustive picture of the resulting proprieties of this generalized model too. Section $5$ is due to a numerical analysis of the capabilities of this extended model: trough a one-to-one comparison among performances of the classical versus the relativistic Hopfield model, we prove how systematically our extension out-performs w.r.t its classical limit and, in particular, we show how the spurious states of the standard Hopfield model are almost entirely pruned by its extension.
\newline
Finally Section $6$ is left for our conclusions and a brief summary of future outlooks.

\section{A teaspoon of neural networks from the statistical mechanics perspective: Boltzmann learning and Hopfield retrieval}
\begin{figure*}[t]
\begin{center}
\includegraphics[width=.60\textwidth]{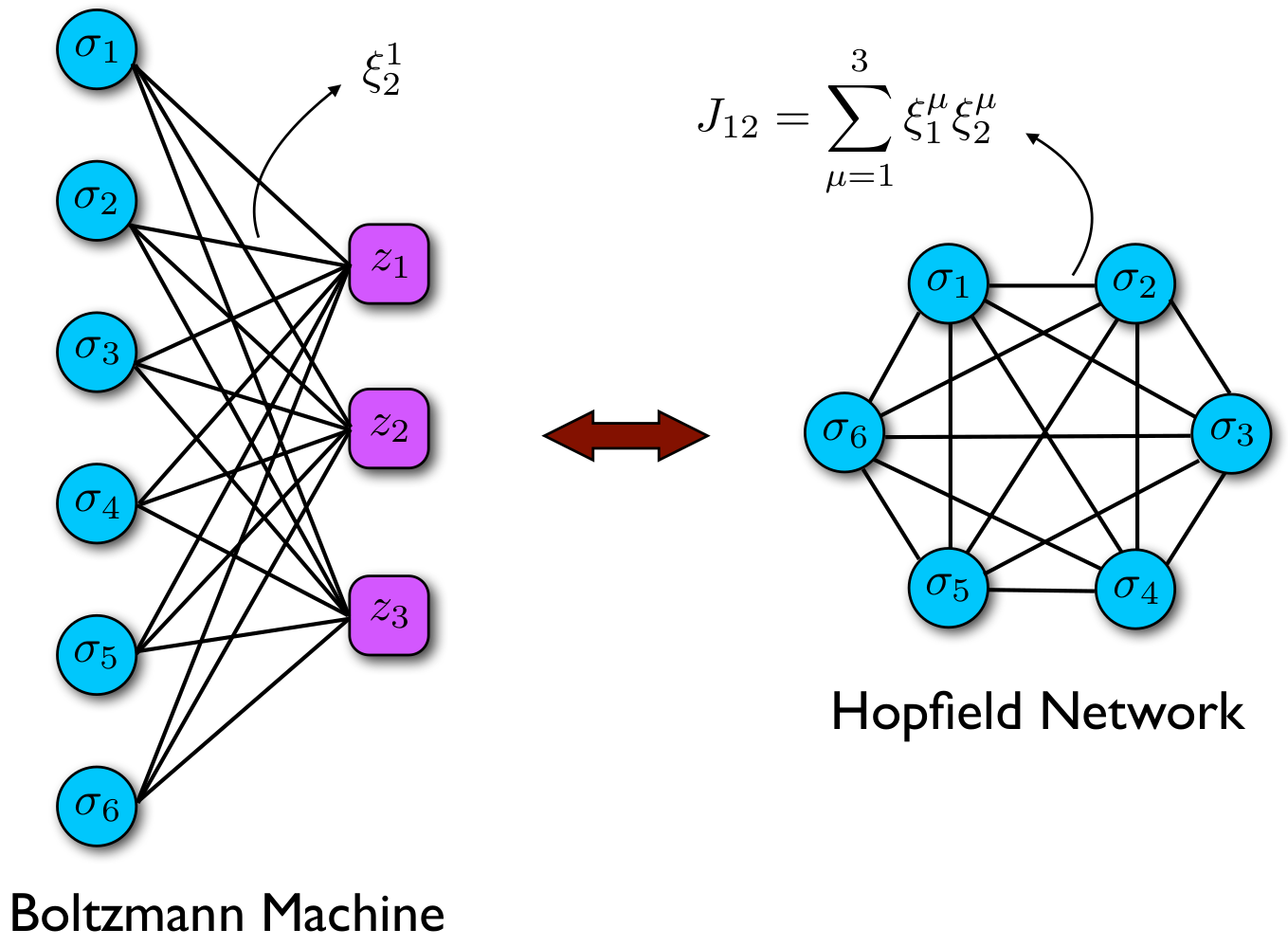}
\caption{{\bfseries Restricted Boltzmann machine  and associative Hopfield network.} Left panel: example of a restricted Boltzmann machine equipped with $6$ visible neurons $\sigma_i,\ i \in (1,...,6)$ in the visible (or input) layer and $3$ hidden units $z_{\mu},\ \mu \in (1,...,3)$ in the hidden layer. The weights connecting them form the $N \times P$ matrix $\xi_i^{\mu}$. Right panel: example of the corresponding Hopfield network, whose six visible neurons $\sigma_i,\ i \in (1,...,6)$ retrieve as patterns stored in the Hebb matrix $J_{ij}=\sum_{\mu}^p \xi_i^{\mu}\xi_j^{\mu}$ the three vectors $\bold{\xi}^{\mu},\ \mu \in (1,...,3)$, each pertaining to a {\em feature}, {\it i.e.} one of the three $z_{\mu}$ hidden variables of the (corresponding) RBM.} \label{fig:equivalenza}
\end{center}
\end{figure*}
For long time, {\em machine learning} (meant as statistical learning of characteristic features -or patterns- from input data \cite{Coolen}) and {\em machine retrieval} (meant as  recognition of previously stored patterns -or features- \cite{Amit}) have been treated separately, the former mainly addressed from a statistical inference perspective, the latter mainly tackled trough statistical mechanics techniques. Hereafter instead, we show how Restricted Boltzmann machines learn by a stochastic dynamics  ({\it e.g.} contrastive divergence \cite{Hinton1}), whose long term relaxation converges to a Gibbs measure of an effective Hamiltonian that turns out to be sharply the Hopfield model for pattern recognition; then, in a streamlined way, we summarize the properties of the latter by means of standard statistical mechanical arguments: this is a preliminary step required to correctly frame the mechanical analogy in the broad world of neural networks.
\newline
In a nutshell, a Boltzmann machine is a two-party system (or a {\em bipartite spin-glass} in the statistical mechanical vocabulary \cite{Agliari-Barattolo,BarraHowGlassy,Barra-bip,Barra-RBMsPriors}) whose parties (or {\em layers} to keep the original jargon \cite{Hinton1,DLbook}) are called {\em visible} -the one that receives the input from the outside world- and {\em hidden} -the one dedicated to figure out correlations in the data presented to the visible layer \cite{BM1,Bengio1}, (see Fig. \ref{fig:equivalenza}, left panel).
\newline
Keeping Fig. \ref{fig:equivalenza} in mind, each layer is composed by spins/neurons  ($N$ for the  visible and $P$ for the hidden) and these spins can be chosen with high generality, ranging from discrete-valued, e.g., we can select Ising neurons for the visible layer $\sigma_i = \pm 1$, $i \in (1,...,N)$, to real-valued, {\it e.g.} we can chose Gaussian neurons for the hidden layer $z_{\mu} = \mathcal{N}[0,1]$, $\mu \in (1,...,P)$ \cite{Barra-RBMsPriors,HintonLast}.
\newline
Analogously, the entries of the weight's matrix $\xi_i^{\mu}$ can be both real or discrete: generally speaking, while continuous weights allow the usage of stronger learning rules  ({\it i.e.} contrastive divergence with simulated annealing \cite{Hinton1,kirkpatrick}) w.r.t. their binary counterparts (whose typical  learning rule is the old fashioned Hebbian one \cite{Amit}) and are thus more convenient during the learning stage, binary weights are more convenient in the retrieval stage, when pattern recognition is in order \cite{Barra-RBMsPriors}.
\newline
This trade-off gave rise to a broad plethora of variations on theme (see for instance \cite{Florent1,Florent3,Rodriguez,Huang1,Huang2,Hinton1,HintonLast} and references therein), whose best setting (for our purposes) lies  in a Boltzmann machine equipped with a Gaussian hidden layer and a Boolean visible layer \cite{Barra-RBMsPriors}: the structure of this typical restricted Boltzmann machine can thus be coded into the following effective cost-function
\be\label{RBM}
H_{\ti N}(\boldsymbol{\sigma},\boldsymbol z|\boldsymbol{\xi})= -\frac{1}{\sqrt{N}}\sum_{i,\mu}^{N,P}\xi_i^{\mu}\sigma_i z_{\mu}-\sum_i^N h_i \sigma_i,
\ee
where the $N$ external fields $h_i$ act as {\em bias terms}  from a statistical inference perspective: from the statistical mechanics viewpoint, these {\em one body} interactions are always trivial and, while playing a key role in machine learning, they will be discarded soon (with no loss of generality  as they can always be re-introduced at any time later).
\newline
Crucially, the dynamics of the weights  ({\it i.e.} the slow variables, also called {\em synapses} in standard neural network jargon \cite{Amit,Coolen}) and the dynamics of the spins  ({\it i.e.} the fast variables, also called {\em neurons} in  standard neural network jargon \cite{Amit,Coolen}) evolve separately, adiabatically on different timescales (or more properly {\em epochs} \cite{DL1}), such that we can average away the fast neural scale when studying the slow synaptic dynamics (learning) and, likewise, we can keep synapses quenched when interested at the fast neural dynamics (retrieval).
\newline
Typically, a set of data vectors is presented to the visible layer of  the machine  ({\it i.e.} the so called {\em training set}) and, under the assumption that these data have been independently generated by the same probability distribution $Q(\boldsymbol{\sigma})$, the ultimate goal of the machine is to make an inner representation of $Q(\boldsymbol{\sigma})$, say $P(\boldsymbol{\sigma})$, that is as close as possible to the original one.
\newline
The way in which $P(\boldsymbol{\sigma})$ approximates $Q(\boldsymbol{\sigma})$ is usually achieved by the minimization of the Kullback-Leibler cross entropy $D(Q,P)$, defined as
\be\label{KL}
D(Q,P) \equiv \sum_{\boldsymbol{\sigma}}Q(\boldsymbol{\sigma})\ln \left( \frac{Q(\boldsymbol{\sigma})}{P(\boldsymbol{\sigma})}\right),
\ee
as, once introduced a small parameter $\epsilon$  ({\it i.e.} the {\em learning rate}), by imposing $\Delta \xi_i^{\mu} = - \epsilon (\partial D(Q,P)/(\partial \xi_i^{\mu}))$ we get 
$$
\Delta D(Q,P) = \sum_{i,\mu} \frac{\partial D(Q,P)}{\partial \xi_i^{\mu}}\Delta \xi_i^{\mu} = - \epsilon \left( \sum_{i,\mu}\left( \frac{\partial D(Q,P)}{\partial \xi_i^{\mu}} \right)^2 \right) \leq 0,
$$
and analogously for the response of the Kullback-Leibler cross entropy to a variation in the biases $h_i$. This is a {\em secure learning rule} thanks to the definite sign in the last term above.
\newline
Weights in the Boltzmann machine are symmetric and this suffices for Detailed Balance to hold \cite{Amit,Coolen}: the latter guarantees that the long term limit of any (non-pathological) stochastic dynamics will always end up in a Gibbs measure, hence the mathematical structure of the probability distribution $P$ is known and this allows generating explicit algorithms, among which the following {\em contrastive divergence} criterion  is probably the most applied \cite{Coolen}:
\be
\frac{\partial D(Q,P)}{\partial \xi_i^{\mu}} = \epsilon   \left(  \langle \sigma_i \sigma_j \rangle_{clamped} - \langle \sigma_i \sigma_j \rangle_{free} \right).
\ee
In this equation, the subscript {\em clamped} means that the averages $\langle . \rangle$ must be evaluated when the visible layer is forced to assume data values, while {\em free} means that the averages are the standard statistical mechanical ones: this recipe is a very elementary yet powerful learning rule as, roughly speaking, it simply tries to make the theoretical correlation functions as close as possible to the empirical ones.\footnote{This argument can be expanded up to arbitrarily $N$-points correlation functions by paying the price of adding extra hidden layer and we believe this way of reasoning to lie at the core of Deep Learning when inspected via sisordered statistical mechanics \cite{DL1,HopfieldNew}.}
\newline
When the machine is able to reproduce the statistics stored in the training data correctly,  the internal weights have been rearranged such that if the machine is now asked to generate vectors according to $P(\boldsymbol{\sigma})$, their statistical properties will coincide with those of the input data: the machine {\em has learnt} a representation of the reality it has been fed with (at least at the lowest orders it has been constrained to infer, {\it e.g.} averages and variances if we deal with pairwise cost-functions).
\newline
Now we question about the retrieval capabilities of such a machine, that is, what kind of features will this machine discover when provided with further data. Since hidden units are independent in RBMs, it is $P(\boldsymbol z|\boldsymbol{\sigma})=\prod_{\mu=1}^P P(z_{\mu}|\boldsymbol{\sigma})$, and, analogously, $P(\boldsymbol{\sigma}| \boldsymbol z) = \prod_{i=1}^N P(\sigma_i|\boldsymbol z)$, hence to get their joint and marginal distributions we can use Bayes Theorem in a very elementary way, {\it i.e.} $P(\boldsymbol{\sigma}, \boldsymbol z)=P(\boldsymbol z|\boldsymbol{\sigma})P(\boldsymbol{\sigma})=P(\boldsymbol{\sigma}|\boldsymbol z)P(\boldsymbol z)$, thus
\begin{eqnarray}
P(\boldsymbol{\sigma},\boldsymbol z) &\propto& \sum_{\{\boldsymbol{\sigma}\}}^{2^N}\int \prod_{\mu=1}^P dz_{\mu} \exp\left( -\frac{1}{2}\sum_{\mu} z_{\mu}^2 + \sum_{i,\mu} \xi_i^{\mu} \sigma_i z_{\mu} + \sum_i h_i \sigma_i \right), \label{hopfield}\\
P(\boldsymbol{\sigma}) &\propto& \sum_{\{\boldsymbol{\sigma}\}}^{2^N} \exp \left( \frac{1}{2}\sum_{ij}^N \left(\sum_{\mu}^P \xi_i^{\mu}\xi_j^{\mu} \right) \sigma_i \sigma_j  + \sum_i h_i \sigma_i \right),\label{trallalero}
\end{eqnarray}
namely, carrying out the integrals over the $z_{\mu}$'s to move from \eqref{hopfield} to \eqref{trallalero}, in the solely visible layer we are left with the Gibbs measure of the following Hopfield Hamiltonian
\be\label{toy-hopfield}
H_N(\boldsymbol{\sigma}\vert\boldsymbol{\xi}, \boldsymbol h) = -\frac{1}{2N} \sum_{i,j=1}^N \left(\sum_{\mu=1}^P \xi_i^{\mu}\xi_j^{\mu} \right) \sigma_i \sigma_j -\sum_{i=1}^N h_i \sigma_i.
\ee
Note that in the above Eq. \eqref{toy-hopfield}  the one-body interactions show how the biases in the RBM automatically translate into the {\em thresholds for firing} in the associative neural networks jargon, where the Hopfield model is mostly used: from now on we set $h_i \equiv 0$ for all the $N$ neurons with no loss of generality.
\newline
To understand the retrieval capabilities of the above cost function from a statistical mechanical perspective, it is useful to introduce the $P$ Mattis magnetizations, defined as
\begin{equation}\label{Mattis}
m_{\mu} \equiv \frac1N \sum_i^N \xi_i^{\mu}\sigma_i, \ \ \ \ \mu \in (1,...,P).
\end{equation}
They cover the pivotal role of {\em order parameters} in Hopfield theory as they capture the relaxation of the network. This can happen in essentially three different ways:
\newline
(a) network's dynamics ends up in a pure state, namely in an attractor that coincides exactly with one of the $P$ stored patterns (such that its corresponding Mattis magnetization becomes one as the neurons are all aligned with the pattern entries in a pure state).
\newline
(b)  network's dynamics ends up  in a spurious state, namely in an unwanted mixture of pure states  (whose amount, spontaneously, grows very quickly as $P$ raises \cite{Amit})  such that its corresponding Mattis magnetization becomes strictly different from zero, albeit sensibly smaller than the one.
\newline
(c)  network's dynamics has a failure and ends up in a random (ergodic) sample of the phase space, where the Mattis parameters are all zero (the network relaxed to a fully noisy state).
\newline
As we can write the Hopfield Hamiltonian via the Mattis order parameters as
$$H_{\ti N}(\boldsymbol{\sigma}\vert \boldsymbol{\xi}) \sim -\frac{1}{2N} \sum_{i,j=1}^N \left(\sum_{\mu=1}^P \xi_i^{\mu}\xi_j^{\mu} \right) \sigma_i \sigma_j = - N \sum_{\mu}^P m_{\mu}^2,$$
by a minimum energy request  ({\it i.e.} minimizing the cost-function) it follows that the candidate minima for the Hopfield model are those states $\bold{\sigma}$ that force one of the $P$ Mattis magnetizations to assume value $1$ or a combination of Mattis magnetization to be different from zero: remarkably, some equilibrium states, the global minima, do coincide with the $P$ patterns (that can thus be retrieved) and we are then tempted to associate the spontaneous relaxation of the network into the attractor to the cognitive process of information retrieval.
\newline
To quantify this statement, the standard route \cite{Amit,Coolen} is to introduce and study the free energy related to the model, that is, to express and extremize it in terms of the Mattis order parameters, as briefly summarized hereafter for the low-storage case, namely when $\lim_{N \to \infty} P/N = 0$.
\newline
The free energy of the Hopfield model, in the low storage regime, reads as
\begin{eqnarray}
\alpha(\beta)&=& \lim_{N \to \infty} \frac{1}{N}\ln \sum_{\{\boldsymbol{\sigma}\}}^{2^N} e^{ \frac{\beta}{2N} \sum_{ij}^N \left(\sum_{\mu}^P \xi_i^{\mu}\xi_j^{\mu} \right) \sigma_i \sigma_j} = \lim_{N \to \infty} \frac{1}{N}\ln \sum_{\{\boldsymbol{\sigma}\}}^{2^N} e^{ \frac{\beta}{2N} \sum_{\mu} \left( \sum_i \xi_i^{\mu}\sigma_i \right)^2 }\\
&=& \label{serpe} \lim_{N \to \infty} \frac{1}{N}\ln \sum_{\{\boldsymbol{\sigma}\}}^{2^N} \int \prod_{\mu}^P \frac{d m^{\mu}}{\sqrt{2\pi}}e^{-N \left(\frac{\beta}{2}\sum_{\mu}m_{\mu}^2 + \sum_{\mu}\sum_i m^{\mu}\xi_i^{\mu}\sigma_i \right)},
\end{eqnarray}
where the first line is the bare definition of the free energy, while in the second line we used the Gaussian integral representation: the exponent in  \eqref{serpe} is extensive due to the $N$ factor multiplying all $O(1)$ terms, thus it can be evaluated with a saddle point argument, whose extremization returns
\begin{eqnarray}\label{sella1}
\alpha(\beta)&=& \ln2 + \langle \log\cosh\big(\beta \sum_{\mu} m^{\mu}\bold{\xi}_{\mu}\big)\rangle_\xi - \frac{\beta}{2}\langle \sum_{\mu}m_{\mu}^2\rangle_{\bold{\xi}},\\ \label{sella2}
\frac{d\alpha(\beta)}{d\langle m_{\mu}\rangle} &=&0 \Rightarrow \langle m_{\mu}\rangle= \langle \bold{\xi}_{\mu}\tanh\big(\beta \sum_{\mu} m^{\mu}\bold{\xi}_{\mu} \big)\rangle_{\bold{\xi}}, \ \ \ \mu \in (1,...,P).
\end{eqnarray}
Eq. \eqref{sella1} is the explicit expression of the free energy of the model in terms of the Mattis magnetizations: the free energy is the entropy minus the cost function (at a given noise level $\beta$), hence extremizing the free energy we obtain the maximum entropy solutions of the cost-function minimization. These are coded in  the last equation, Eq. \eqref{sella2}, that is a {\em self-consistent} equation for the order parameter and quantifies the strength of the retrieval (note that we have one self-consistent equation for any of the $P$ patterns).
\newline
Whatever the route, that is starting from the Hebbian prescription \cite{Hebb} as in the original Hopfield paper \cite{Hopfield} or upon marginalization over the hidden layer in Boltzmann machine (hence after learning via contrastive divergence) \cite{BarraEquivalenceRBMeAHN}, unfortunately, we always end up in a network whose attractors are by far more than the stored patterns (namely more than the solely {\em pure states} we would see retrieved by the network) \cite{Amit}, the excess stock of (local) minima being constituted by the so-called {\em spurious states}: the simplest example is a 3-pattern mixture defined as follows
\be\label{statospurio3}
\sigma_i = \text{sign} \left(\xi_i^1+\xi_i^2+\xi_i^3\right).
\ee
The Mattis overlap of this state with any of the three patterns is -for large networks- $m^{\nu}= N^{-1}\sum_i \xi_i^{\nu}\sigma_i = 0.5$ (for $\nu=(1,2,3)$) hence, while smaller in amplitude than the Mattis overlap of a pure state (whose amplitude is one), it is still a meta-stable state: if orbiting in the surrounding, the network can be attracted by such spurious states and converge to them rather than to the pure ones. Unfortunately, as the patterns are added linearly to the memory kernel, there is an exponential (combinatorial) proliferation of these unwanted meta-stable states in the retrieval landscape of the Hopfield network such that Hopfield himself suggested a procedure to prune -or remove- (a part of) them from his coupling matrix \cite{HopfieldUnlearning}. In a nutshell, Hopfield's idea is again transparent, elegant and brilliant: as there are sensibly much more spurious states ({\it i.e.} metastable minima) than pure states (global minima), let's start the system at random and make a quench  ({\it e.g.} a search for minima with steepest descent rather than conjugate gradient or stochastic algorithms). With high probability, in this way, the system will end up in a spurious state: we can collect this equilibrium configuration - called $\langle \sigma_i \sigma_j \rangle_{spurious}$ hereafter - and subtract it from the memory kernel, via
$$
J_{ij} = \sum_{\mu=1}^P \xi_i^{\mu}\xi_j^{\mu} - \langle \sigma_i \sigma_j \rangle_{spurious}.
$$
We can do this iteratively and check that effectively the network becomes progressively cleared from these nasty attractors: this procedure is called {\em unlearning} \cite{unlearning0,unlearning1,unlearning2,unlearning3} and it has been linked to REM sleep \cite{Crick} (offering a possible intriguing picture for its interpretation) due to the effectiveness of the random starting point setting for the quenching procedure in consolidating memories (phenomenon to be eventually correlated with the rapid eye movements in that part of sleep).

\section{A mechanical formulation of the (classical) Hopfield networks}

Aim of this section is to recover formulae \eqref{sella1} and \eqref{sella2} without relying any longer on the standard statistical mechanical guidance, that is, we will no longer use
nor the Maximum Entropy neither the Minimum Energy principles  ({\it i.e.} overall the standard free energy extremization). Indeed, by observing that the free energy plays as an action in a suitable mechanical analogy,
we can  import an arsenal of mathematical tools to  investigate its properties (originally developed within the framework of analytical mechanics).
In particular, we will show that the free energy -as any  proper action- obeys an Hamilton-Jacobi PDE, whose solution returns Eq. \eqref{sella1}: the variational principle for the free energy minimization will be
expressed, in this context, by the Least Action Principle (and will return Eq. \eqref{sella2}).

\subsection{Preliminary definitions regarding the (classical) Hopfield free energy}

In this section we introduce more formally all the quantities and observables we will deal with, starting by the first
\begin{Definition} The Hopfield model is described by the Hamiltonian
\be\label{2.1}
H_\ti{N} (\boldsymbol\sigma\vert\boldsymbol{\xi})= -\sum _{i< j}^N J_{ij} \spin{i} \spin{j} ,
\ee
where $\spin{i}=\pm 1$, $i \in (1,...,N)$ are Ising spins (or  McCulloch$\&$Pitts neurons), while the coupling matrix $J$ is defined in terms of $P$ patterns  $\bold{\xi}^{\mu}$, $\mu \in (1,...,P)$ as follows \begin{eqnarray}\label{2.2}
J_{ij} &=& \frac{1}{N}\sum_{\mu=1}^p \pat{\mu}{i} \pat{\mu}{j} = \frac{1}{N} \boldsymbol \xi_{i} \cdot \boldsymbol \xi_{j},\\
\frac12 &=& P(\xi_i^{\mu}=+1)=P(\xi_i^{\mu}=-1).
\end{eqnarray}
\end{Definition}
Using the label $\beta \in \mathbb{R}^+$ to tune the (inverse) level of noise in the network, the partition function of the model is then introduced as
\be\label{2.3}
Z_\ti{N}(\beta) =\sum _{\{\boldsymbol\sigma\}}^{2^N}\exp \left({-\beta H_\ti N (\boldsymbol\sigma,\boldsymbol{\xi})}\right)\sim\sum _{\{\boldsymbol\sigma\}}^{2^N}\exp
\left(\frac{\beta}{2} \sum _{i,j=1}^N J_{ij} \spin{i} \spin{j} \right),
\ee
where $\sim$ means an equality up to a  constant (due to self-interactions $i=j$) that is negligible in the thermodynamic limit, as long as we keep the load of the system sub-linear in the volume ({\it i.e.} $P \neq cN$, for some positive constant $c$), i.e., in the low storage scenario under study.
\begin{Definition} The (intensive) free-energy $\alpha(\beta)$ is defined \footnote{Note that we use here and everywhere the subscript $N$ when we work at fixed volume $N$, its lacking expressing quantities already evaluated in the thermodynamic limit.} as
\be\label{2.5}
\alpha (\beta)= \lim _{N\rightarrow\infty} \alpha _\ti{N}(\beta), \ \ \ \alpha _\ti{N}(\beta) = \frac{1}{N}\log Z_\ti{N}(\beta), \\
\ee
\end{Definition}
Considering a generic function of the weights and the neurons $\Phi(\boldsymbol{\sigma}, \boldsymbol{\xi})$, we introduce also the average over the patterns as
\be\label{2.7}
\langle \Phi(\boldsymbol{\sigma}, \boldsymbol{\xi})\rangle_{{\ti\ensuremath{\xi}}}= \lim_{N\rightarrow \infty}\frac{1}{N}\sum _{i=1} ^N \Phi(\boldsymbol{\sigma}, \boldsymbol{\xi}_i),
\ee
(where $\boldsymbol\xi_{i} = (\pat{1}{i},\dots ,\pat{p}{i})$), and the standard Boltzmann averages, i.e.
\be\label{2.4}
\langle \Phi(\boldsymbol{\sigma}, \boldsymbol{\xi}) \rangle  =\frac{\sum _{\{\boldsymbol\sigma\}}\Phi(\boldsymbol{\sigma}, \boldsymbol{\xi})\exp \left({-\beta H_\ti N
(\boldsymbol\sigma\vert \boldsymbol \xi)}\right)}{Z_\ti{N}(\beta) }.
\ee

\par\medskip

Now we start to generalize the former observables in order to make them {\em flexible enough} for our mechanical construction. To this task we introduce $P$ {\em spatial} variables $x_{\mu}\in \mathbb{R}$, $\mu \in (1,...,P)$ and $1$ temporal variable $t \in \mathbb{R}^+$ that we use to generalize the partition function \eqref{2.3} as follows:
\be\label{2.8}
Z_\ti{N}(t,\bold{x})= \sum _{\{\boldsymbol\sigma\}}\exp \left(-\frac{t}{2N} \sum _{i,j=1}^N \boldsymbol\xi_{i} \cdot \boldsymbol\xi_{j}  \spin{i} \spin{j} + \sum _{i=1}^N
\bold{x} \cdot \boldsymbol\xi_{i} \spin{i}  \right).
\ee
We call $\langle \cdot \rangle _{t,\bold{x}}$ the expectation value of any quantity of the network with this new partition function.
\newline
Note that, crucially, taking $t=-\beta$ and $\bold{x}= \bold{0}$, this generalized partition function (Eq. \eqref{2.8}) reduces to the standard one of statistical mechanics (Eq. \ref{2.3}).
\newline
The next observable will play as the generalized free energy of the Hopfield model from a statistical mechanical perspective, while covers the role of an action in the present mechanical analogy.
\begin{Definition}
By using the partition function \eqref{2.8}, we introduce the interpolating free energy $\alpha_N(t, \bold{x})$ as
\be\label{2.9}
\alpha_{\ti N}(t, \bold{x}) = \frac1N \ln Z_\ti{N}(t,\bold{x})= \frac1N \ln \sum _{\{\boldsymbol\sigma\}}\exp \left(-\frac{t N}{2} \bold{m}_{\ti N}(\boldsymbol\sigma)^2 + N \bold{x}\cdot \bold{m}_{\ti N}(\boldsymbol\sigma) \right),
\ee
where
\be\label{2.10}
\bold{m}_{\ti N}(\boldsymbol\sigma)= \frac{1}{N}\sum _{i=1}^N \boldsymbol\xi_{i} \spin{i},
\ee
is the vector whose components are the Mattis magnetizations $m_{\mu}, \ \mu \in (1,...,P)$.
\end{Definition}
Finally, it is elementary to check that the space-time derivatives of the interpolating free energy \eqref{2.9} read as
\be\label{2.11}
\begin{split}
\frac{\partial 	\alpha _{\ti N} (t,\bold{x})}{\partial t}&= -\frac{1}{2}\langle\bold m_{\ti N}^2\rangle_{t,\bold{x}},\\
\frac{\partial 	\alpha _{\ti N} (t,\bold{x})}{\partial x^\mu }&=\langle m_{\ti N}^\mu\rangle_{t,\bold{x}}.
\end{split}
\ee
This is the starting point for the mechanical analogy.

\subsection{Construction of the (classical) Hopfield free energy via Hamilton-Jacobi PDE}

The mechanical analogy we pursue lies in the Hamilton-Jacobi framework as we show in this section.
\begin{Proposition}
By contruction, $\alpha_{\ti N}(t,\bold{x})$ obeys the following (classical) Hamilton-Jacobi PDE and it plays de facto as an  action in this mechanical analogy
\begin{eqnarray}\label{2.22}
&& \frac{\partial \alpha_\ti N}{\partial t}+\frac{1}{2}\left(\frac{\partial\alpha_\ti N}{\partial x^\mu}\right)^2+V_{\ti N} (t,\bold{x})=0,\\
&& V_\ti{N} (t,\bold{x})=\frac{1}{2}\left(\langle \bold m_{\ti N}^2\rangle_{\ti\ensuremath{t,\bold{x}}}\, - \langle \bold{m}_{\ti N}\rangle_{\ti\ensuremath{t,\bold{x}}}^2\,\right)=\frac{1}{2}\sum_{\mu=1}^P\left(\langle m_{\ti N,\mu}^2\rangle_{\ti\ensuremath{t,\bold{x}}}\, - \langle {m}_{\ti N,\mu}\rangle_{\ti\ensuremath{t,\bold{x}}}^2\,\right).
\end{eqnarray}
This partial differential equation describes -even at finite volume $N$- the motion of a classical (non relativistic) particle, with unitary mass\footnote{Note that, according to equation \eqref{2.22}, the classical momentum is $\langle \bold{m} \rangle _{\ti\ensuremath{t,\bold{x}}}$.} in $P+1$ dimensions.
\end{Proposition}
\begin{Remark}\label{rem:1}
In the thermodynamic limit, away from critical point, the Mattis magnetizations self-average, i.e.
\be
\lim_{N\rightarrow \infty} (\langle \bold m^2_{\ti N} \rangle _{\ti\ensuremath{t,\bold{x}}}-\langle \bold{m}_{\ti N} \rangle _{\ti\ensuremath{t,\bold{x}}}^2)=0,
\ee
such that $\lim_{N \to \infty} V_{\ti N}(t, \bold{x}) = V(t, \bold{x}) = 0$.
\end{Remark}
\begin{Remark}
In the thermodynamic limit, the motion has space-time symmetries whose No$\ddot{e}$ther currents, derived respectively for the momentum conservation and for the energy conservation as
\begin{eqnarray}
\lim_{N\rightarrow \infty} (\langle \bold m_{\ti N}^2 \rangle _{\ti\ensuremath{t,\bold{x}}}-\langle \bold{m}_{\ti N} \rangle _{\ti\ensuremath{t,\bold{x}}}^2)&=&0 \ \ \ \ \textit{Momentum Conservation}\\
\lim_{N\rightarrow \infty} (\langle \bold m_{\ti N}^4 \rangle _{\ti\ensuremath{t,\bold{x}}}-\langle \bold{m}_{\ti N}^2 \rangle _{\ti\ensuremath{t,\bold{x}}}^2)&=&0 \ \ \ \ \textit{Energy Conservation},
\end{eqnarray}
mirror the classical self-averaging properties in the statistical mechanical jargon.
\end{Remark}
As a consequence, in the thermodynamic limit, the particle's motion of the mechanical analogy paints a Galilean trajectory, i.e., a straight line $\bold{x}= \bold{x}_0 +\langle \bold{m}\rangle_{\ti\ensuremath{t,\bold{x}}}\cdot(t-t_0) $.
\newline
Further, in this limit, the determination of an explicit expression of the free energy in terms of the Mattis magnetizations reduces to the explicit calculation of the action of this free motion. As Cauchy conditions we choose $t_0=0, \bold{x}_0$ (note that the choice $t_0=0$ kills the coupling between the spins) such that
\be\label{2.23}
\alpha(t,\bold{x}) = \alpha(0,\bold{x}_0)+\int_0 ^t dt'\mathcal L(t'),
\ee
where $\mathcal L= \frac{1}{2}\langle \bold{m}\rangle_{\ti\ensuremath{t,\bold{x}}}^2$ is the Lagrangian. The latter is a constant of motion as $V(t, \bold{x})=0$, hence the solely calculations required are due to evaluate the (trivial) Cauchy condition
\be\label{2.24}
\alpha(0,\bold{x}_0) =\log 2 +\langle \log \cosh \bold{x_0} \cdot \bold{\xi} \rangle_{\xi}.
\ee
Merging the two observations and writing $\bold{x}_0 = \bold{x}(t) - \langle \bold{m}\rangle_{\ti\ensuremath{t,\bold{x}}} t$, we have
\begin{Theorem}
The infinite volume limit of the Hopfield action \eqref{2.9} reads as
\be\label{2.25}
\alpha (t,\bold{x})= \log 2+\langle \log \cosh (\bold{x} - t \langle \bold{m}\rangle_{\ti\ensuremath{t,\bold{x}}}) \cdot \boldsymbol{\xi}\rangle_{\bold{\xi}}+\frac{t}{2} \langle
\bold{m}\rangle_{\ti\ensuremath{t,\bold{x}}}^2.
\ee
\end{Theorem}
\begin{Remark}
Note that by setting $t=-\beta$ and $\bold{x}=\bold{0}$, the (classical) Hopfield free energy is obtained (check Eq. \eqref{sella1}).
\end{Remark}
\begin{Corollary}
Within the mechanical analogy, the expression \eqref{2.25} is subjected to the Last Action Principle $\delta \alpha (t,\bold{x}) =0$.
\end{Corollary}
Indeed we can perform an infinitesimal variation $\langle m^{\mu}\rangle_{t, \boldmath{x}} \to \langle m^{\mu}\rangle_{t, \bold{x}} + \delta \langle m^{\mu}\rangle_{t, \bold{x}}$ and verify that
\be
\delta \alpha (t,\bold{x}) = \frac{\partial \alpha (t,\bold{x})}{\partial \langle m^{\mu}\rangle_{t, \bold{x}}}\delta \langle m^{\mu}\rangle_{t, \bold{x}} =
\tanh(x - \langle m^{\mu}\rangle_{t, \bold{x}} t)(-t \delta \langle m^{\mu}\rangle_{t, \bold{x}}) + \langle m^{\mu}\rangle_{t, \bold{x}} \delta \langle m^{\mu}\rangle_{t, \bold{x}} t = 0,
\ee
automatically implies the self consistency equations \eqref{sella2} to hold.

\section{A mechanical formulation of the (relativistic) Hopfield networks}

The mechanical analogy can now be used to note pathologies in previous treatment that somehow shine in its transparent calculations and will led us to a very natural extension of the Hopfield cost-function: actually there are two main observations waiting to be done. The first is that, as the free energy plays as an action, we can think at the exponent in the Maxwell-Boltzmann weight  ({\it e.g.} as expressed in the partition function \eqref{2.8}) as the product of the $P+1$ momentum-energy tensor with the $P+1$ metric tensor  ({\it i.e.} $-t \cdot E + \bold{x}\cdot \bold{m})$, and observe that the underlying metrics is not Euclidean, rather Minkowskian, as in special relativity. The second is that in classical mechanics the velocity is unbounded, while here the magnetization  is obviously bounded by one, {\it i.e.} $\langle m^{\mu} \rangle \leq c \equiv 1$: the whole points straight to a natural {\em relativistic generalization}, that, indeed, is the content of this Section.

\subsection{Preliminary definitions regarding the (relativistic) Hopfield free energy}

Because in the mechanical analogy the Hopfield cost-function \eqref{2.1} reads as the kinetic energy associated to the fictitious particle, a natural extension of Hopfield model is constituted by its relativistic deformation, i.e.
\be\label{3.1}
-\frac{\bold{m}_{\ti N}^2}{2}\rightarrow - \sqrt{1 +\bold{m}_{\ti N}^2}.
\ee
This calls for the next
\begin{Definition}
The Hamiltonian for the relativistic Hopfield model is defined as
\be\label{relativistico}
H_{\ti N}(\boldsymbol{\sigma}\vert \boldsymbol{\xi}) = - N \sqrt{1 + \bold{m}_{\ti N} (\boldsymbol{\sigma})^2}.
\ee
\end{Definition}
Two observations are in order here: at first note that Taylor-expanding the Hamiltonian \eqref{3.1} in the Mattis magnetizations we collect a series of many-body (P-spin) contributions (hence in the direction suggested by Hopfield recently in regards of Deep Learning \cite{HopfieldNew}), as
\be\label{eq:3.4}
-\frac{H_\ti N(\boldsymbol{\sigma}\vert\boldsymbol{\xi})}{N} = 1+\frac{1}{2N^2}\sum_{ij}(\boldsymbol\xi_{i} \cdot  \boldsymbol\xi_{j})\spin i\spin j-\frac{1}{8 N^4}\sum _{ijkl} (\boldsymbol\xi_{i}\cdot
\boldsymbol\xi_{j} )(\boldsymbol\xi_{k} \cdot\boldsymbol\xi_{l})\spin i \spin j\spin k \spin l+\dots
\ee
further, the r.h.s. of Eq. \eqref{eq:3.4} is an alternate-sign series, hence it will have both contributions in learning (those with the minus sign) and in unlearning (those with the plus sign \cite{unlearning0,unlearning1,unlearning2}), with the prevailing contribution played by standard Hebbian learning (as the series converges). We will show in Section $5$ that the beneficial role of unlearning lies in destabilizing the retrieval of spurious memories, thus resulting in enhanced network's performances w.r.t. the classical limit  ({\it i.e.} the standard Hopfield model).
\newline
Once this new model is introduced, we extend partition function, free energy and mechanical analogy properly from the previous sections, e.g.
\begin{Definition}
The {\em relativistic} partition function and free energy, related to Hamiltonian \eqref{relativistico}, read as
\begin{eqnarray}\label{3.2}
Z_\ti{N} (\beta) &=& \sum _{\{\boldsymbol\sigma\}}^{2^N}\exp \left({\beta N}\sqrt{1+ \bold{m}_{\ti N}(\boldsymbol\sigma)^2}  \right),\\ \label{azzozzo}
\alpha(\beta) &=& \lim_{N \to \infty} \alpha_{\ti N}(\beta), \ \ \ \alpha_{\ti N}(\beta) = \frac1N \ln Z_\ti{N} (\beta).
\end{eqnarray}
\end{Definition}
As we did in the previous section, we now read the relativistic partition function from a purely mechanical perspective.
\begin{Definition}
The generalized partition function and the action of the relativistic Hopfield model, suitable for the mechanical analogy, read as
\begin{eqnarray}\label{3.4}
Z_\ti{N} (t,\bold{x}) &=& \sum _{\{\boldsymbol\sigma\}}\exp \left(N(-{t}\sqrt{1+ \bold{m}_{\ti N}(\boldsymbol\sigma)^2} +  \bold{x}\cdot \bold{m}_{\ti N}(\boldsymbol\sigma)
)\right), \\ \label{FER}
\alpha_{\ti N}(t,\bold{x}) &=& \frac1N \ln Z_\ti{N} (t,\bold{x}).
\end{eqnarray}
\end{Definition}
Note the minus sign in front of the {\em spatial part} of the the partition function: this is the Minkowskian signature $(+,-,\dots,-)$, useful to represent the exponent in a Lorentzian form as, using the Einstein notation, $-N x_\ti A p^\ti A=-N\eta_{\ti A \ti B} x^{\ti A}p^{\ti
B}$ with $x^\ti{A}=(t,\bold{x})$, $p^\ti A(\boldsymbol\sigma) = (\sqrt{1+ \bold{m}_{\ti N}(\boldsymbol\sigma)^2}, \bold{m}_{\ti N}(\boldsymbol\sigma))$ and $\eta_{\ti A \ti
B}=\text{diag}(+,-,\dots,-)$ is the standard flat metric of Minkowski spacetime, $A,B=0,1,\dots,p$.
\newline
The expectation values $\langle \dots \rangle$ and $\langle \dots \rangle _\bold{\xi}$ are obviously defined in
the same way as \eqref{2.4} and \eqref{2.7}, but with respect to this new partition function \eqref{3.4}.
\newline
Finally, when setting $t=- \beta$ and $\bold{x}=\bold 0$ we obtain the {\em relativistic statistical mechanical framework} we are interested in.
\par
Again, we start our relativistic generalization of the mechanical analogy by computing the space-time derivatives of the free energy. The relevant ones are
\be\label{3.5}
\begin{split}
\frac{\partial \alpha _\ti{N}(t,\bold{x})}{\partial t}&= -\langle\sqrt{1+\bold{m}_{\ti N}^2} \rangle_{t,\bold{x}}, \\
\frac{\partial \alpha _\ti{N}(t,\bold{x})}{\partial x^\mu}&= \langle{m_{\ti N}^\mu} \rangle_{t,\bold{x}}, \\
\frac{\partial ^2\alpha _\ti{N}(t,\bold{x})}{\partial t^2}&= N (\langle 1+ \bold{m}_{\ti N}^2\rangle_{t,\bold{x}} -  \langle\sqrt{1+\bold{m}_{\ti N}^2} \rangle _{t,\bold{x}}^2),\\
\nabla ^2\alpha _\ti{N}(t,\bold{x}) &= N (\langle\bold m_{\ti N}^2 \rangle_{t,\bold{x}} -  \langle \bold{m}_{\ti N} \rangle _{t,\bold{x}}^2).
\end{split}
\ee

\subsection{Construction of the (relativistic) Hopfield free energy via Hamilton-Jacobi PDE}

\begin{Proposition}
By construction $\alpha_{\ti N}(t,\bold{x})$ obeys the following (relativistic) Hamilton-Jacobi PDE
\be\label{3.6}
\partial_t^2\,\alpha _\ti{N}-\nabla^2\,\alpha _\ti{N}=N\left(1-\left({\partial_t \alpha _\ti{N}}\right)^2+\left(\nabla \alpha_\ti N\right)^2\right),
\ee
or in the manifestly covariant form
\be\label{3.7}
\left(\partial _{\ti A} \alpha _\ti N\right)^2 +\frac{1}{N}\Box \,\alpha _\ti N= 1,
\ee
where $\Box$ represents the D'Alambertian, i.e.,  $\Box = \partial _\ti A \partial ^\ti A$ (still in Einstein notation).
\end{Proposition}
\begin{Remark}
Requiring that the derivatives of the action are regular functions in $x^{\mu}, t$, in the thermodynamic limit, a.s.\footnote{{\em Almost surely} because when ergodicity breaks down a gradient's catastrophe prevents regularity even in the infinite volume limit \cite{Barra-quantum}.}
\be\label{3.8}
\left(\partial _{\ti A}\alpha  \right)^2  = 1.
\ee
\end{Remark}
From the mechanical perspective, in the thermodynamic limit, the $P+1$-momentum of the particle reads as
\be\label{3.9}
p^{\ti A} =-\frac{\partial \alpha}{\partial x_{\ti A}}=(\langle \sqrt{1+ \bold{m}^2}\rangle_{t,\bold{x}},\langle  \bold{m} \rangle _{t,\bold{x}}).
\ee
In terms of this momentum, the equation for the action \eqref{3.8} is, using the relativistic mechanics language, the on-shell relation \cite{Cabibbo} for a free particle with unitary mass.

Note that the fictitious particle of this mechanical analogy moves on the straight lines $\bold{x} =\bold{x}_0 +\bold{v}(t-t_0)$ for arbitrary $(t_0,\bold{x}_0)$, where the velocity $\bold{v} $ is related to the spatial part of the momentum as $\langle m \rangle_{t,\bold{x}} = \gamma \bold{v}$ and $\gamma$ is the usual Lorentz factor. In other words:
\be\label{3.10}
\bold{v} = \frac{\langle\bold{m} \rangle _{t,\bold{x}}}{\sqrt{1+\langle\bold{m} \rangle _{t,\bold{x}}^2}}\quad\Leftrightarrow \quad \langle\bold{m} \rangle _{t,\bold{x}}= \frac{\bold{v}}{\sqrt{1 -\bold{v}^2}}.
\ee
The Lorentz factor can be therefore written as $\gamma = \sqrt{1+\langle\bold m\rangle _{t,\bold x}^2} $. Summing all these observations together, we end up in the determination of an explicit expression for the relativistic free energy in terms of the Mattis magnetizations as it reduces to the calculation of the action of this free motion. As Cauchy conditions we still keep $t_0=0, \bold{x}_0$ (where, again, $t_0=0$ kills the coupling between the spins), such that
\be\label{3.11}
\begin{split}
\alpha (t,\bold x)&= \alpha (0,\bold x_0)+\int ^t _0 dt'\mathcal L(t')=\\
&=\alpha(0,\bold x_0)-\frac{t}{\gamma}=\alpha (0,\bold x_0)-\frac{t}{\sqrt{1+\langle\bold m\rangle _{t,\bold x}^2} },
\end{split}
\ee
since the Lagrangian $\mathcal L = -\gamma^{-1}$ is constant on the classical trajectories.  Thus, again we have
\be\label{3.12}
\alpha (0,\bold x_0)=\log 2+\langle \log \cosh \bold x_0 \cdot \boldsymbol{\xi}\rangle_{{\xi}}.
\ee
Noting that $\bold x_0 = \bold x- \bold v t$ with \eqref{3.10}, and setting $t=-\beta$, $\bold x= \bold 0$ (in order to reobtain the statistical mechanical framework), we can write the next
\begin{Theorem}
The free energy density of the relativistic Hopfield network in the thermodynamic limit reads as
\be\label{3.13}
\alpha (\beta)=\log 2+\langle \log \cosh \left(  \beta\, \boldsymbol{\xi} \cdot\frac{\langle\bold m \rangle }{\sqrt{1+\langle\bold m \rangle ^2}}\right)
\rangle_{{\xi}}+
\frac{\beta}{\sqrt{1+\langle\bold m\rangle ^2} }.
\ee
\end{Theorem}
In the usual statistical mechanical settings, now by imposing thermodynamic prescriptions (namely requiring the free energy to be extremal with respect to the order parameters), we would get the following self-consistence conditions
\be\label{3.14}
\frac{d \alpha (\beta)}{d \langle m ^{\mu} \rangle} = 0 \Rightarrow \langle m ^{\mu} \rangle = \langle \bold{\xi} ^\mu \tanh \left( \beta\, \boldsymbol{\xi} \cdot\frac{\langle\bold m \rangle }{\sqrt{1+\langle\bold m \rangle ^2}}\right)
\rangle _{\bold{\xi}}.
\ee
\begin{Remark}
Note that Eqs. \eqref{3.14} are exactly those prescribed by the Least Action Principle $\delta \alpha(t, \bold{x})=0$ as
\begin{eqnarray}
\delta \alpha(t, \bold{x}) &=& \frac{\partial \alpha(t, \bold{x})}{\partial \langle m^{\mu}\rangle_{t,\bold{x}}}\delta \langle m^{\mu}\rangle_{t,\bold{x}}= \\ \nonumber
&=& \tanh\left(x^\mu- \frac{\langle m^{\mu}\rangle_{t,\bold{x}} t}{\sqrt{1+ \langle \bold m\rangle_{t,\bold{x}}^2}}\right)\left( \frac{-t \langle m^{\mu}\rangle_{t,\bold{x}}\delta \langle m^{\mu}\rangle_{t,\bold{x}}}{(1+\langle \bold m\rangle_{t,\bold{x}}^2)^{\frac{3}{2}}}\right) + \frac{\langle m^{\mu}\rangle_{t,\bold{x}} t \delta \langle m^{\mu}\rangle_{t,\bold{x}}}{(1+\langle \bold m \rangle_{t,\bold{x}}^2)^{\frac{3}{2}}}=0.
\end{eqnarray}
\end{Remark}
\begin{Remark}
Note that, if we take the low momentum limit $\vert\langle \bold
m\rangle\vert\ll1$, we can expand the relativistic model at the lowest order
\be\label{3.15}
\begin{split}
\frac{1}{\sqrt{1+\langle\bold m\rangle ^2} }&= 1-\frac{\langle \bold m\rangle ^2}{2}+\mathcal O \left(\langle \bold m \rangle ^3\right),\\
\frac{\bold m}{\sqrt{1+\langle\bold m\rangle ^2} }&=\bold m +\mathcal O \left(\langle \bold m \rangle ^3\right),
\end{split}
\ee
so to recover the classical Hopfield model and results. 
\end{Remark}

\section{Numerical Simulations of the classical and relativistic Hopfield networks}

Once we have the Mattis self-consistent equations \eqref{3.14} we can solve them recursively and obtain the theoretical expectations for the quality of the retrieval of the various pure and spurious states  ({\it i.e.} the intensities of the various Mattis magnetizations), or we can use them to compare Monte Carlo simulations to network's performances. In order to test the capabilities of the relativistic Hopfield network, we compare its results to those of the classical counterpart: to this task we performed different kind of extensive simulations as follows.

\subsection{Stochastic neural dynamics and network's attractors}

In order to test the amplitude and stability of pure and spurious attractors, we performed the following series of numerical simulation: once introduced random numbers $\eta_i$ uniformly sampled from $[-1,+1]$, we set up the following stochastic neural dynamics,
\begin{eqnarray}
\sigma_i(t+1) &=& \text{ sign}\left[\tanh\left(\beta h_i(\boldsymbol{\sigma}(t))\right)+\eta_i(t)\right]\\
h_i(\boldsymbol{\sigma}(t)) &=& \boldsymbol{\xi} \cdot \frac{\langle \bold{m}_{\ti N}\rangle}{\sqrt{1+\langle \bold{m}_{\ti N}\rangle^2}},
\end{eqnarray}
or, in its probabilistic formulation for the entire network,
\begin{eqnarray}\label{markov}
p_{t+1}(\boldsymbol{\sigma})&=&\prod_{i=1}^N \frac12 \left[ 1 + \sigma_i \tanh\left(\beta h_i(\boldsymbol{\sigma}(t))\right)\right]=
\prod_{i=1}^N \frac{e^{\sigma_i h_i(\boldsymbol{\sigma}(t))}}{2 \cosh\left( \beta h_i(\boldsymbol{\sigma}(t)) \right)},\\ \label{enfatizza}
p_{t+1}(\boldsymbol{\sigma})&=& \sum_{\sigma'}W[\boldsymbol{\sigma}'\to \boldsymbol{\sigma}]p_t(\boldsymbol{\sigma}'), \ \ \ \ W[\boldsymbol{\sigma}'\to \boldsymbol{\sigma}]=\prod_{i=1}^{N} \frac{e^{\sigma_i h_i(\boldsymbol{\sigma}(t))}}{2 \cosh\left( \beta h_i(\boldsymbol{\sigma}(t)) \right)}.
\end{eqnarray}
Note that, also in the dynamical evolution of the system, $\beta$ tunes the level of noise in the network because it triggers the amplitude of the hyperbolic tangent (containing the signal): at each time step $t$, for each neuron $\sigma_i$, the field $h_i(t)$ is computed and multiplied by $\beta$, then the hyperbolic tangent is applied to $\beta h_i(t)$ and this number is compared to $\eta_i(t)$: for $\beta \to \infty$ -large signal limit- the hyperbolic tangent becomes a $\pm 1$ sign function thus the random numbers play no longer any role and the dynamics becomes deterministic minimization, at contrary for $\beta \to 0$ the hyperbolic tangent returns zero whatever the field and the dynamics becomes fully random.
\newline
Note further that, in Eq. \eqref{enfatizza}, we emphasized the transition rates $W[\boldsymbol{\sigma}'\to \boldsymbol{\sigma}]$ in order to show their equivalence with the acceptance rates in the Monte Carlo route: indeed, due to the symmetry of the couplings, Detailed Balance ensure that
\begin{equation}\label{DetBal}
p_{\infty}(\boldsymbol{\sigma}) W[\boldsymbol{\sigma}' \to \boldsymbol{\sigma}]=p_{\infty}(\boldsymbol{\sigma}') W[\boldsymbol{\sigma} \to \boldsymbol{\sigma}'],
\end{equation}
hence, remembering that $p_{\infty}$ must assume the maximum entropy Gibbs-expression $p_{\infty} \propto \exp(-\beta H(\boldsymbol\sigma,\boldsymbol{\xi}))$, we have
$$
p_t(\sigma_i \to \sigma_i') =  \frac{1}{1+ e^{\beta[H(\boldsymbol{\sigma}\vert\boldsymbol{\xi})-H(\boldsymbol{\sigma}'\vert\boldsymbol{\xi})]}},
$$
namely the acceptance criterion of the Glauber algorithm used in the successive bulk of simulations (see Sec. \ref{5.2}).
\newline
We use the above dynamics with two different starting assumption as follows:

\begin{figure}[b!]
	\begin{center}
		\centering
		\includegraphics[width=\textwidth]{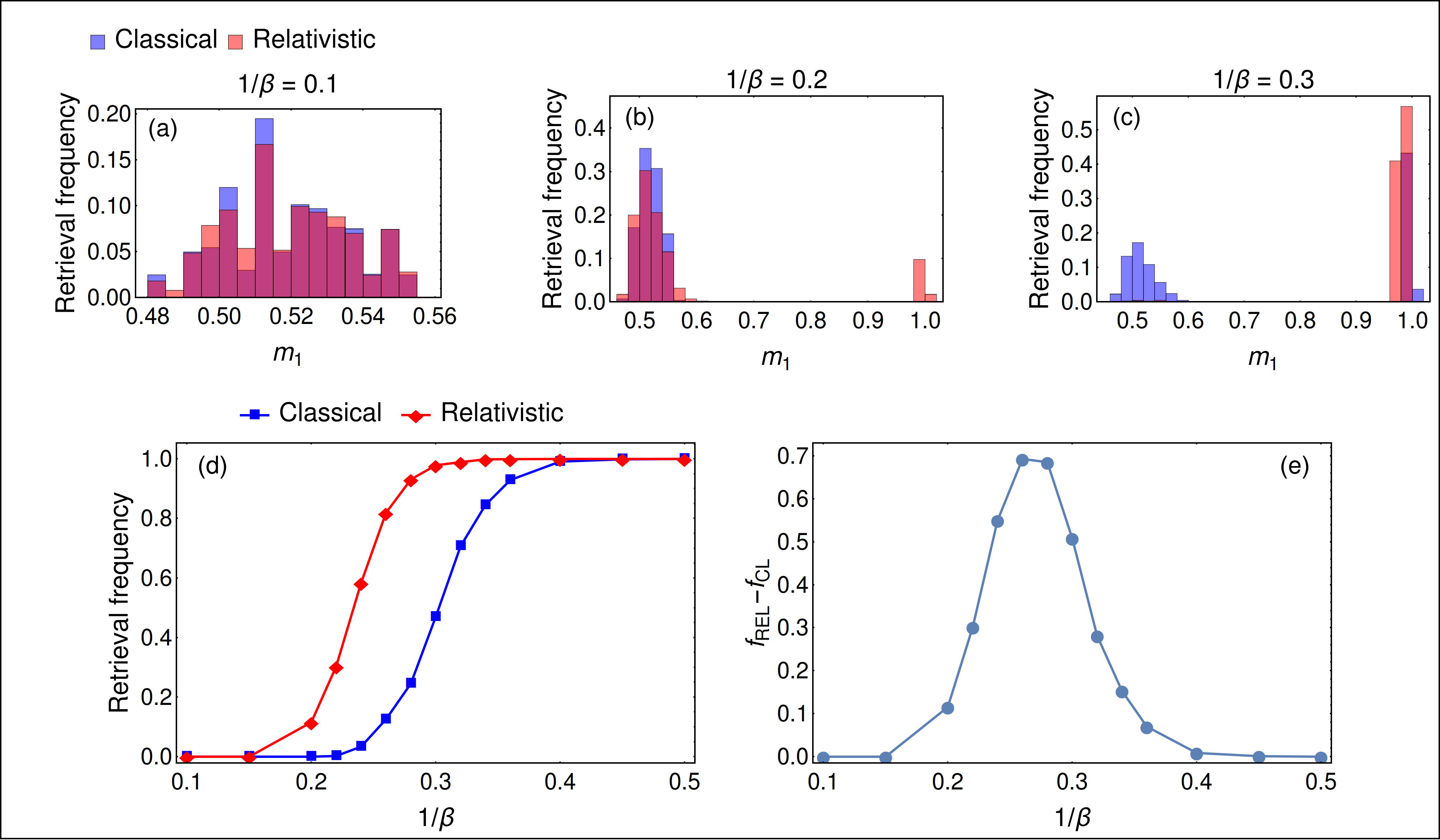}	
	\end{center}
	\caption{{\bfseries First row:  histograms of the maximum Mattis magnetization starting from a spurious state}.  Results for the maximum Mattis magnetization $m_1$ for both classical (blue bars) and relativistic (red bars) Hopfield models for different noise values: $\beta^{-1}=0.1$ (a), $0.2$ (b) and $0.3$ (c). Purple bars are the superposition of blue and red ones. The network whose thermalization is under study is built of by $N=1000$ neurons and $P=3$ stored patterns. Simulations consist in 8000 different runs, organized in 20 different random pattern configurations, 20 randomly selected initial conditions and 20 different stochastic evolutions for each noise level.
		\newline
		{\bfseries Second row: Retrieval performances for spurious states initial conditions.} Results for the retrieval performances in the noise range $0.1 \le \beta^{-1} \le 0.5$. The curves in (d) refer to the number of final states for the classical (blue squares) and the relativistic (red diamonds) models: starting from a spurious state, for lowest thermal energy supply  ({\it e.g.} $\beta^{-1}<0.15$), the spurious states are stable. However, for $\beta^{-1}>0.15$ the relativistic model starts to retrieve properly and at $\beta^{-1} \sim 0.25$ roughly half of the runs end up in a proper pure state. At contrary, for the classical model, we have to reach $\beta^{-1}>0.22$ before obtaining a correct thermalization into a pure state and $\beta^{-1} \sim 0.32$ before half of the runs end up correctly. Mattis magnetizations whose magnitude is larger that an arbitrary threshold (set at $m_{threshold}=0.8$) are coupled to final patterns recognized as ``retrieved'': we choose $m_{threshold} = 0.8$ since it best splits spurious state intensities (whose range is close to $m_{spurious} \sim 0.5$)  and pure state intensities (whose magnitude is close to one). Plot (e) is obtained subtracting the classical curve from the corresponding relativistic one, giving an empirical estimation about how the latter improves the retrieval performances.}\label{spuriousstatestability}
\end{figure}
\begin{figure}[htb!]
	\begin{center}
		\centering
		\includegraphics[width=\textwidth]{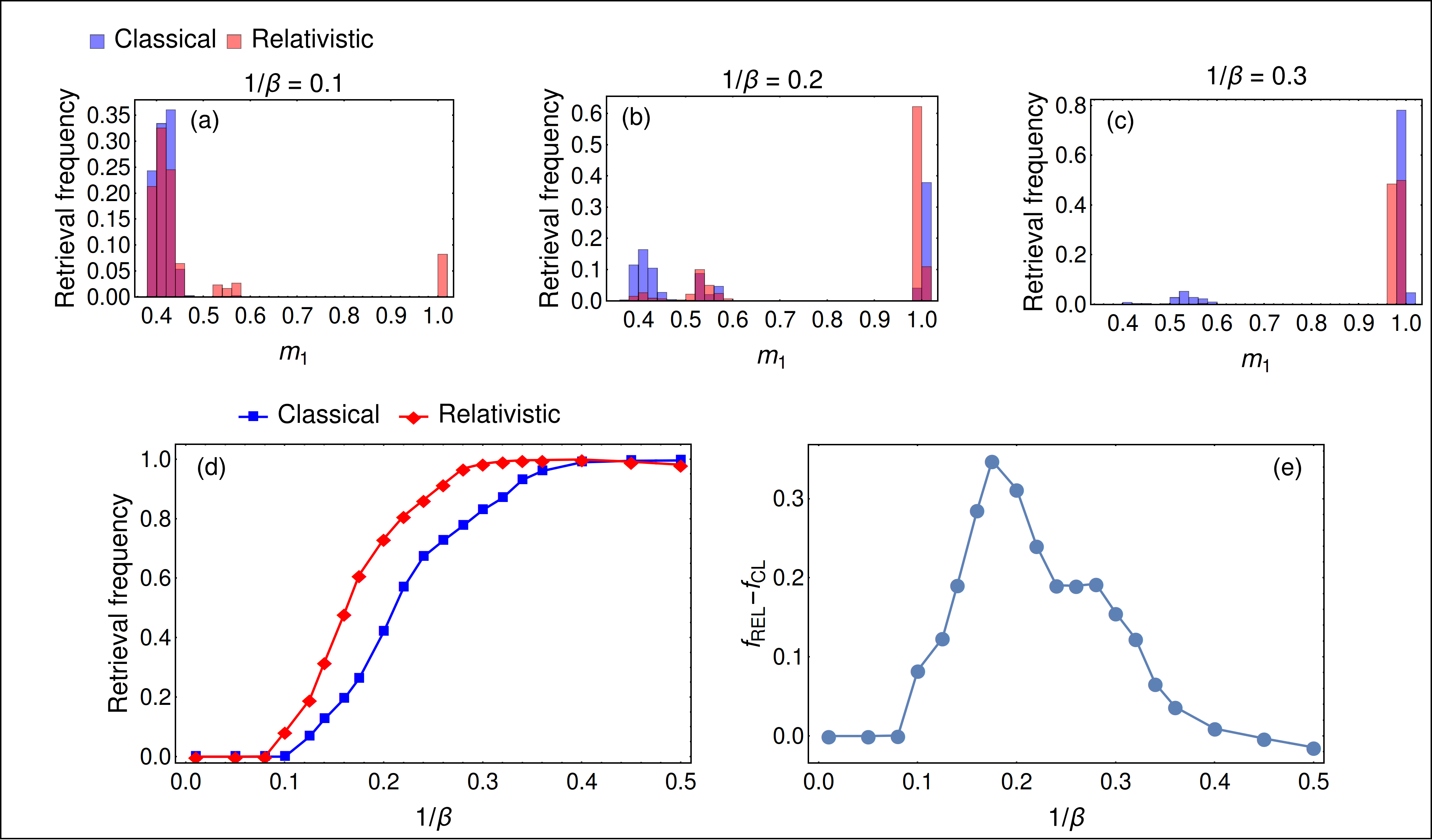}	
	\end{center}
	\caption{{\bfseries First row:  histograms of the maximum Mattis magnetization starting from a spurious state for $P=5$}.  Results for the maximum Mattis magnetization $m_1$ for both classical (blue bars) and relativistic (red bars) Hopfield models for different noise values: $\beta^{-1}=0.1$ (a), $0.2$ (b) and $0.3$ (c). Purple bars are the superposition of blue and red ones. The network whose thermalization is under study is built of by $N=1000$ neurons and $P=5$ stored patterns. Simulations consist in 8000 different runs, organized in 20 different random pattern configurations, 20 randomly selected initial conditions and 20 different stochastic evolutions for each noise level.
		\newline
		{\bfseries Second row: Retrieval performances for spurious states initial conditions.} Results for the retrieval performances in the noise range $0.1 \le \beta^{-1} \le 0.5$. The curves in (d) refer to the number of final states for the classical (blue squares) and the relativistic (red diamonds) models: starting from a spurious state, for lowest thermal energy supply  ({\it e.g.} $\beta^{-1}<0.15$), the spurious states are stable. However, for $\beta^{-1}>0.15$ the relativistic model starts to retrieve properly and at $\beta^{-1} \sim 0.25$ roughly half of the runs end up in a proper pure state. At contrary, for the classical model, we have to reach $\beta^{-1}>0.22$ before obtaining a correct thermalization into a pure state and $\beta^{-1} \sim 0.32$ before half of the runs end up correctly. Mattis magnetizations whose magnitude is larger that an arbitrary threshold (set at $m_{threshold}=0.8$) are coupled to final patterns recognized as ``retrieved'': we choose $m_{threshold} = 0.8$ since it best splits spurious state intensities (whose range is close to $m_{spurious} \sim 0.5$)  and pure state intensities (whose magnitude is close to one). Plot (e) is obtained subtracting the classical curve from the corresponding relativistic one, giving an empirical estimation about how the latter improves the retrieval performances.}\label{spuriousstatestabilityP5}
\end{figure}

\begin{figure}[htb!]
	\begin{center}
		\centering
		\includegraphics[width=\textwidth]{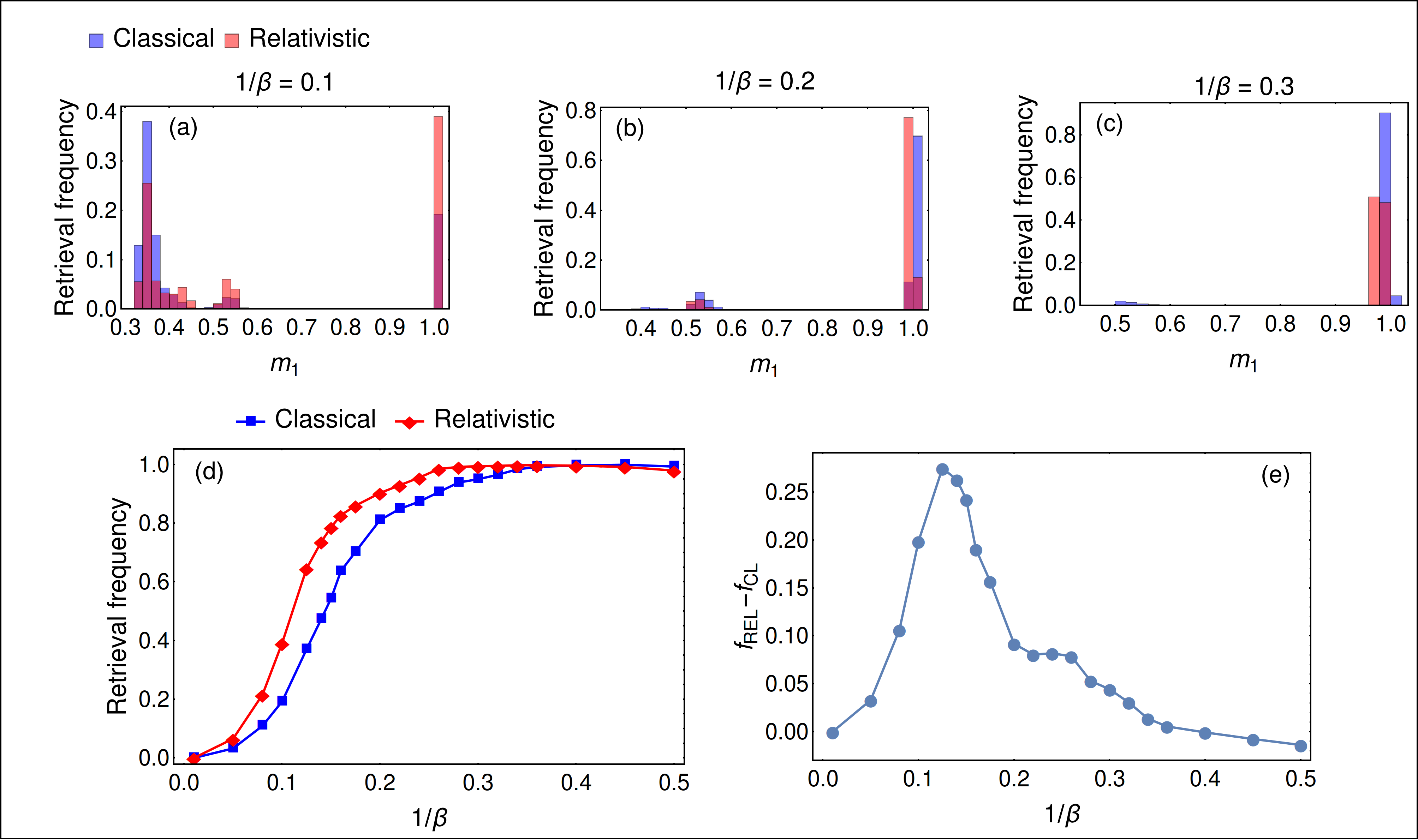}	
	\end{center}
	\caption{{\bfseries First row:  histograms of the maximum Mattis magnetization starting from a spurious state for $P=7$}.  Results for the maximum Mattis magnetization $m_1$ for both classical (blue bars) and relativistic (red bars) Hopfield models for different noise values: $\beta^{-1}=0.1$ (a), $0.2$ (b) and $0.3$ (c). Purple bars are the superposition of blue and red ones. The network whose thermalization is under study is built of by $N=1000$ neurons and $P=7$ stored patterns. Simulations consist in 8000 different runs, organized in 20 different random pattern configurations, 20 randomly selected initial conditions and 20 different stochastic evolutions for each noise level.
		\newline
		{\bfseries Second row: Retrieval performances for spurious states initial conditions.} Results for the retrieval performances in the noise range $0.1 \le \beta^{-1} \le 0.5$. The curves in (d) refer to the number of final states for the classical (blue squares) and the relativistic (red diamonds) models: starting from a spurious state, for lowest thermal energy supply  ({\it e.g.} $\beta^{-1}<0.15$), the spurious states are stable. However, for $\beta^{-1}>0.15$ the relativistic model starts to retrieve properly and at $\beta^{-1} \sim 0.25$ roughly half of the runs end up in a proper pure state. At contrary, for the classical model, we have to reach $\beta^{-1}>0.22$ before obtaining a correct thermalization into a pure state and $\beta^{-1} \sim 0.32$ before half of the runs end up correctly. Mattis magnetizations whose magnitude is larger that an arbitrary threshold (set at $m_{threshold}=0.8$) are coupled to final patterns recognized as ``retrieved'': we choose $m_{threshold} = 0.8$ since it best splits spurious state intensities (whose range is close to $m_{spurious} \sim 0.5$)  and pure state intensities (whose magnitude is close to one). Plot (e) is obtained subtracting the classical curve from the corresponding relativistic one, giving an empirical estimation about how the latter improves the retrieval performances.}\label{spuriousstatestabilityP7}
\end{figure}

\begin{figure}[t!]
	\begin{center}
		\centering
		\includegraphics[width=0.8\textwidth]{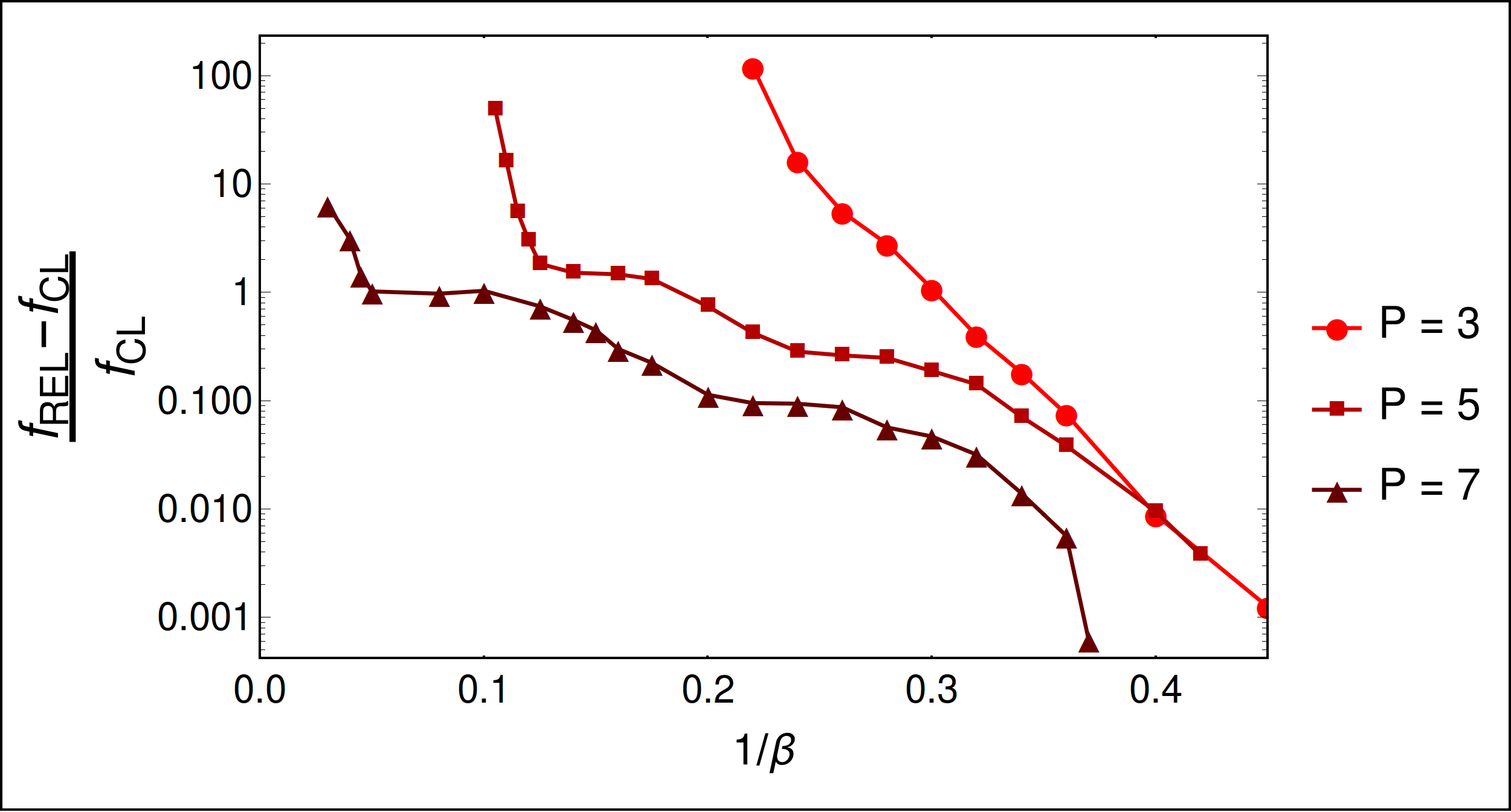}	
	\end{center}
	\caption{{\textbf{Relative improvement as a function of the temperature.}} The plot show the empirical curves for the empirical relative performances $(f_{\text{REL}}-f_{\text{CL}})/f_{\text{CL}}$ of relativistic Hopfield model versus its classical counterpart for $P=3,5,7$, where $f$ is the retrieval frequency of a given model (recall that we say that a pattern is retrieved if the associated Mattis magnetization is higher than a fixed value $m_{threshold}=0.8$). The starting initial condition is aligned to the maximal spurious state (obtained by a majority rule applied to the sum of all patterns, {\it i.e.} $\sigma_{0,i}= \text{sign}(\sum_{\mu} \xi^\mu _i)$ for each $i$). Increasing the value of $P$, such spurious attractors becomes unstable also for the classical model, therefore causing the fast fall-off of the the relative improvement for increasing $P$. For the particular value $P=3$, the curve falls also outside the range and, for low temperature values ($\beta^{-1}\sim 0.1$), the relative improvement is very high {\it i.e.} $(f_{\text{REL}}-f_{\text{CL}})/f_{\text{CL}}\sim 120$, since in this case the retrieval frequency for the relativistic model is finite, while for the classical one is almost zero.}\label{pdependence}
\end{figure}

\begin{itemize}

\item Attractors from spurious initial conditions. We start the system sharply within a spurious state (the 3-patterns mixture for $P=3$, see Eq. \eqref{statospurio3})  then we let it thermalize at a given noise level $\beta^{-1}$ under the Markov dynamics \eqref{markov} and we collect the final state of the relaxation process (for both for the classical and the relativistic models), whose existence is ensured by Detailed Balance \eqref{DetBal}. Results are shown in Fig. \ref{spuriousstatestability} for $P=3$, Fig. \ref{spuriousstatestabilityP5} and Fig. \ref{spuriousstatestabilityP7} for the same analysis respectively for $P=5$ and $P=7$. In the first rows we show the histograms  counting how many times the system ended up in a pure state at three different noise levels $\beta^{-1}= 0.1,\ 0.2,\ 0.3$. In the second rows we plot the frequency ({\it i.e.} the percentage) by which a pure state is reached during this thermalization as a function of the noise level $\beta^{-1}$. Finally, in Fig. \ref{pdependence} we plot the relative improvement of the relativistic extension with respect to Hopfield model for the values $P=3,5,7$.
\newline
Two remarks are in order here: the first is that systematically the relativistic model out-performs w.r.t. its classical counterpart. The second is that such an improvement lies in the range of noise level $0 \leq \beta^{-1} \leq 0.5$, namely exactly in the noise region where spurious states possess an own stability: for noise levels beyond that threshold spurious states are no longer stable and, likewise, there is generally no more reward in the relativistic extension.
\begin{figure}[t!]
	\begin{center}
		\centering
		\includegraphics[width=\textwidth]{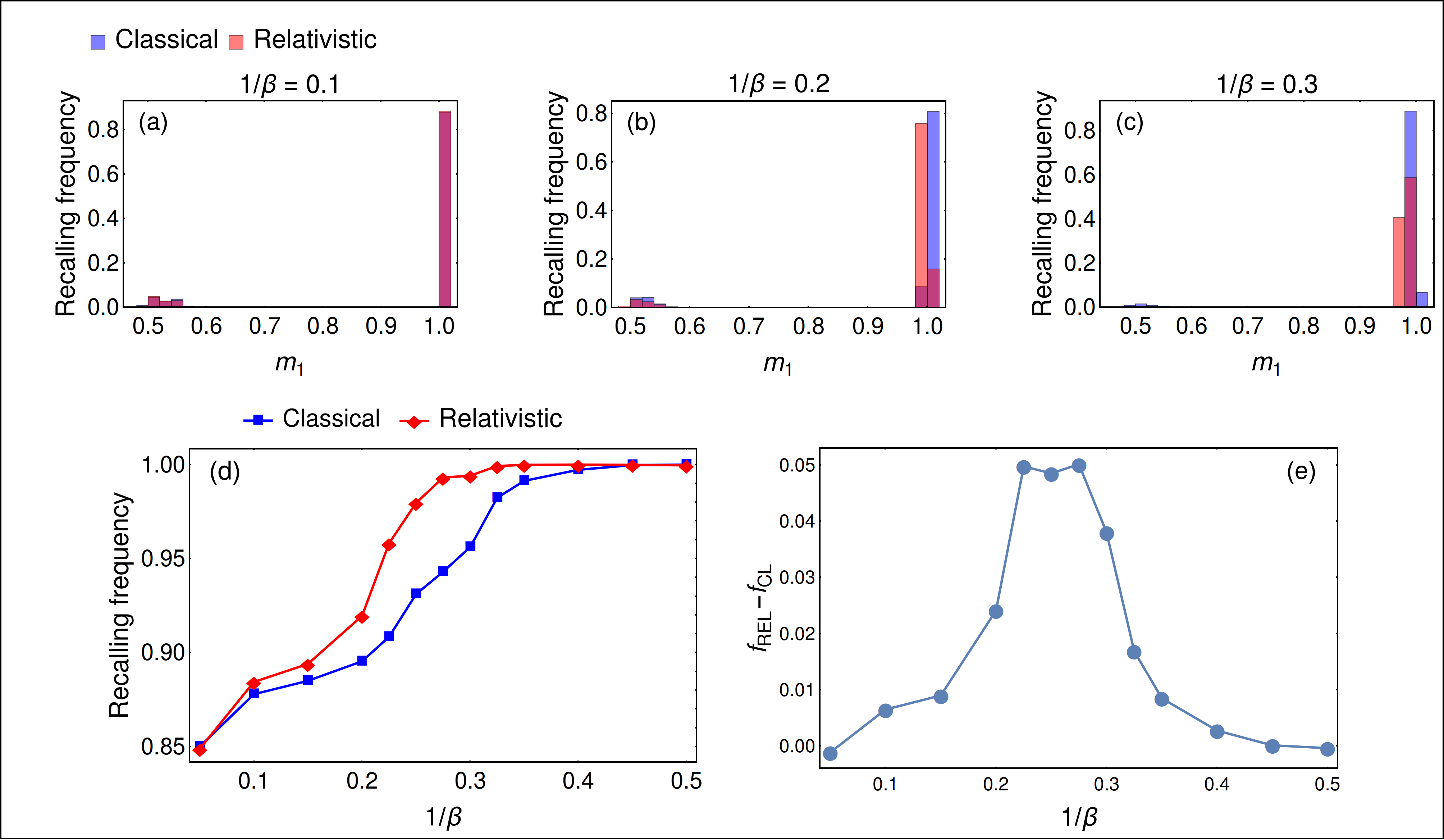}	
	\end{center}
	\caption{{\bfseries First row: histograms of the Mattis magnetization from random initial conditions for $P=3$.} Results for the maximum Mattis magnetization $m_1$ for both classical (blue bars) and relativistic (red bars) Hopfield models for different noise values: $\beta^{-1}=0.1$ (a), $0.2$ (b) and $0.3$ (c). Purple bars are the superposition of blue and red ones. The network whose thermalization is under study is built of by $N=1000$ neurons and $P=3$ stored patterns. MonteCarlo simulations consist in 8000 different runs, organized in 20 different random pattern configurations, 20 randomly selected initial conditions and 20 different stochastic evolutions for each noise level.
		\newline
		{\bfseries Second row: retrieval performances for random initial conditions.} Results for the retrieval performances in the noise range $0.05 \le \beta^{-1} \le 0.5$. The curves in (d) refer to the number of final states for the classical (blue squares) and the relativistic (red diamonds) models: Mattis magnetizations whose magnitude is larger that an arbitrary threshold (set at $m_{trheshold}=0.8$) are coupled to final patterns recognized as ``retrieved'': we choose $m_{threshold} = 0.8$ since it best splits spurious state intensities (whose range is close to $m_{spurious} \sim 0.5$)  and pure state intensities (whose magnitude is close to one). Plot (e) is obtained subtracting the classical curve from the corresponding relativistic one, giving an empirical estimation about how the latter improves the retrieval performances.}\label{purestatestability}
\end{figure}
\item Attractors from random initial conditions. We prepared the system within a fully random initial configuration, then we let it thermalize at a given noise level $\beta^{-1}$ under the Markov dynamics \eqref{markov} and we collect the final state of the relaxation process (for both for the classical and the relativistic models), whose existence is ensured by Detailed Balance \eqref{DetBal}. Results are shown in Fig. \ref{purestatestability}.
Again, in the first row we show the histograms  counting the times the system ended up in a pure state at three different noise levels $\beta^{-1}= 0.1,\ 0.2,\ 0.3$. In the second row we plot the frequency ({\it i.e.} the percentage) by which a pure state is reached during this thermalization as a function of the noise level $\beta^{-1}$.
\newline
Again we highlight that, systematically, the relativistic model shows increased performances that its classical counterpart (as long as the spurious states are locally stable).

\subsection{Comparison between Monte Carlo simulations and analytical outcomes}\label{5.2}
\begin{figure}[htb!]
	\begin{center}
		\centering
		\includegraphics[width=\textwidth]{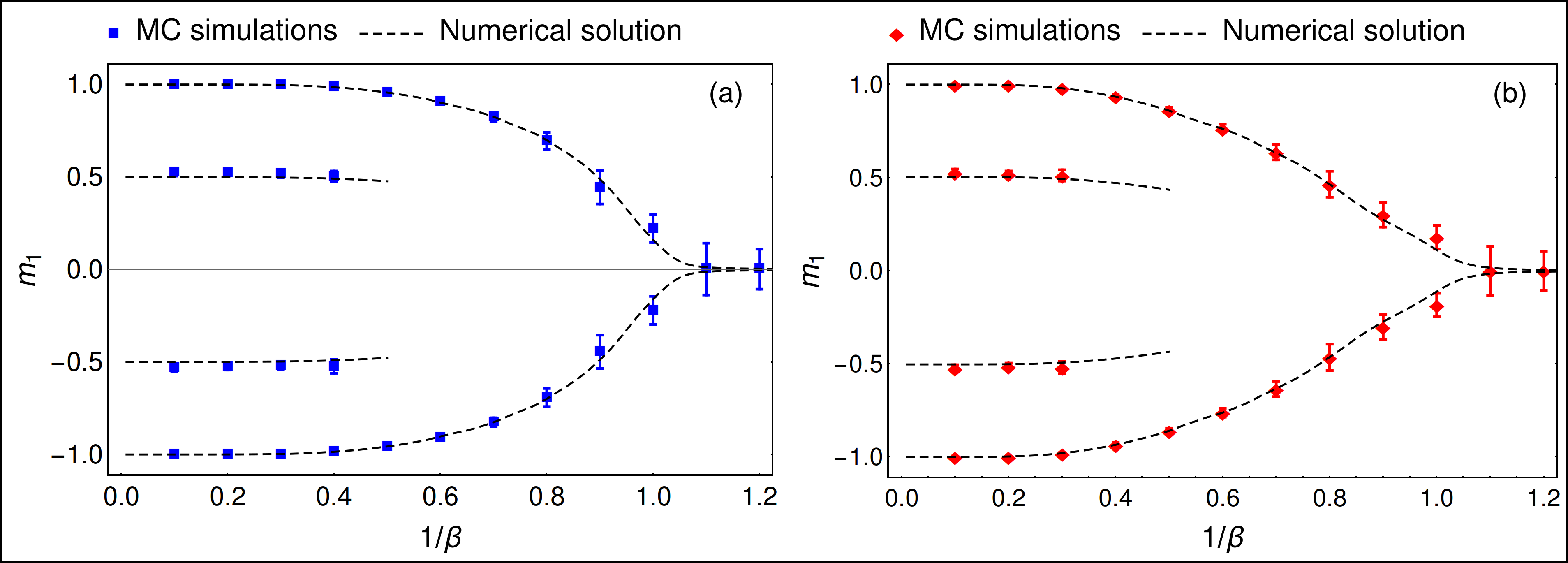}	
	\end{center}
	\caption{{\bfseries MC simulations VS self-consistency prediction for the Mattis order parameter.} Results on the evolution of $\langle m^{\mu}\rangle$ vs $\beta$ are presented for MC simulations of the classical (a, blue squares) and relativistic (b, red diamonds) models for a network with $N=1000$, where $P=3$ (orthogonal)  patterns are stored. The data points are averages over $20$ different pattern realizations, for each of which we sampled $20$ random initial conditions ({\it i.e.} $8000$ runs at any given noise level). MC results are compared to the main branches (pure states) of the theoretical solution (dashed black curves) of the self-consistency Eqs. \eqref{2.11} and \eqref{3.14}). Note the presence of the spurious states (evidenced by the two segments of magnetization's values $m_1 \sim \pm 0.5$) for noise level $\beta^{-1}\leq 0.45$ and note further that the variances of the data-points spread at the bifurcation point ({\it i.e.} at $\beta^{-1}\sim 1$), as typical in a second order phase transition.}\label{selfs}
\end{figure}
At first we performed extensive Monte Carlo runs to check how the evolution of the Mattis magnetizations (versus the noise level) predicted analytically coincides with the numerical one.

\begin{figure}[htb!]
	\begin{center}
		\centering
		\includegraphics[width=\textwidth]{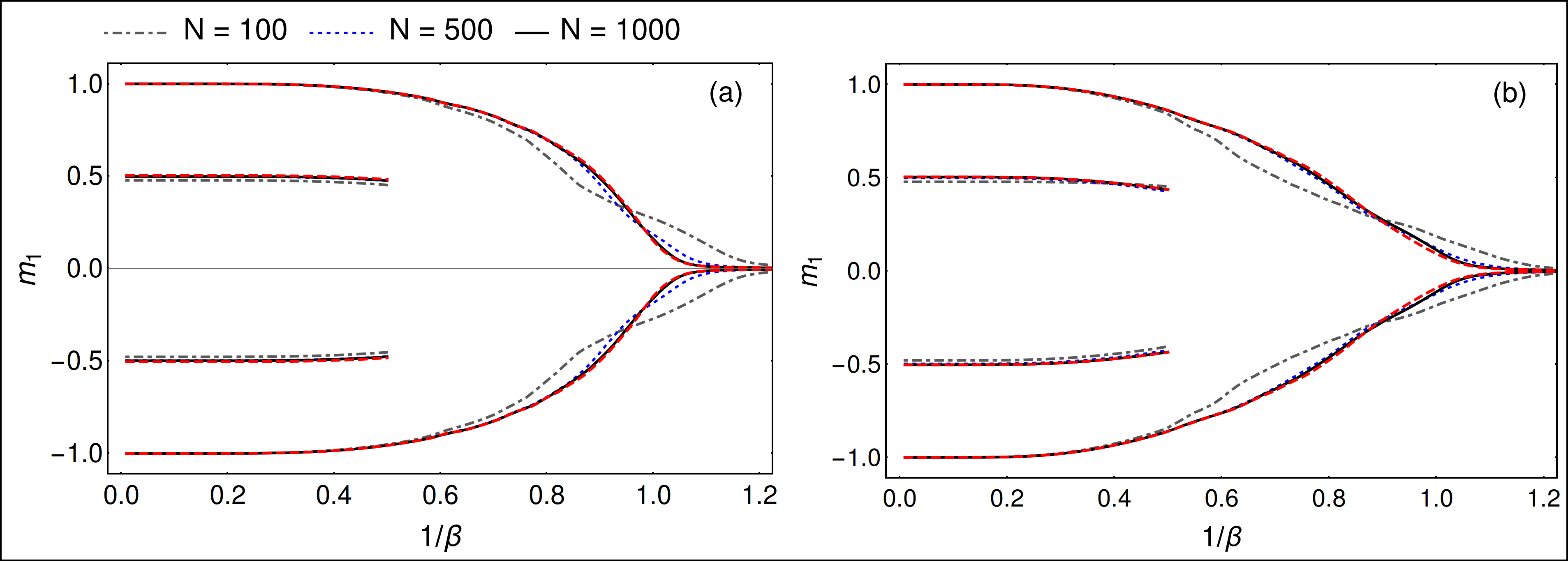}	
	\end{center}
	\caption{{\bfseries Finite size dependence.} Results for different networks of size $N=100,\ 500,\ 1000,\ 2000$ are shown for the classical Hopfield model (left panel) and for the relativistic extension (right panel): as expected, as $N$ grows, the curves approach the analytic solution (obtained in the thermodynamic limit) while for the smallest values of $N$  ({\it e.g.} $N=100$) there is no trace of the phase transition as expected.}\label{finitesize}
\end{figure}
Monte Carlo simulations have been implemented iterating a Glauber dynamics within the following scheme:
\begin{itemize}

\item Select a neuron at random and compute the difference  $\Delta H(\boldsymbol{\sigma},\boldsymbol{\xi})$ in the cost function due to its spin flip.

\item If $\Delta H(\boldsymbol{\sigma}\vert\boldsymbol{\xi}) <0$ (hence the flip is convenient), the move is accepted with probability $\propto \exp(\beta \Delta H(\boldsymbol{\sigma}\vert\boldsymbol{\xi}))/[1-\exp(\Delta H(\boldsymbol{\sigma}\vert\boldsymbol{\xi}))]$, otherwise is rejected (Glauber criterion).

\item If $\Delta H(\boldsymbol\sigma\vert\boldsymbol{\xi}) >0$ (hence the flip is not convenient), the move is rejected.
\end{itemize}
Again due to Detailed Balance, Hopfield's Hamiltonian plays also as a Lyapounov function \cite{Coolen} hence (properly tuning the noise level $\beta$) we can control its convergence to a minimum (a global one -pure state- or a local one -spurious state) whose attainment is reached when the spins share the same sign of the fields acting on them (and the simulation then stops as any spin flip can only raise the energy value).
\newline
In Fig. \ref{selfs} we show the behavior of the Mattis order parameter as a function of the noise in the network both as predicted by the theory, using the self-consistent Eq. \eqref{2.11} for the classical Hopfield model -left panel- and the self-consistent Eq. \eqref{3.14} for the relativistic counterpart -right panel-, as well as the same behaviour as obtained from simulations: these have been obtained via standard Monte Carlo runs for a system built up by $N=1000$ neurons (whose details are reported in the caption of Fig. \ref{selfs}).
\newline
Despite redundant, we stress that two branches -rather than one- are reported in Fig. \ref{selfs} due to the unbroken model's spin-flip symmetry: both the {\em pattern} $\bold{\xi}^1$ and the {\em anti-pattern} $-\bold{\xi}^1$ are attractors for the neural dynamics here but clearly this does not happen when the thresholds for firing $h_i$ are re-introduced in the theory (see Eq. \eqref{toy-hopfield}).
\newline
In Fig. \ref{finitesize} we compare the solution at various sizes of the network: as it shines from the plots, both for the classic case (left panel) as well as for the relativistic generalization (right panel), as $N$ grows the analytical scaling is approached.



\begin{figure}[b!]
	\begin{center}
		\centering
		\includegraphics[width=\textwidth]{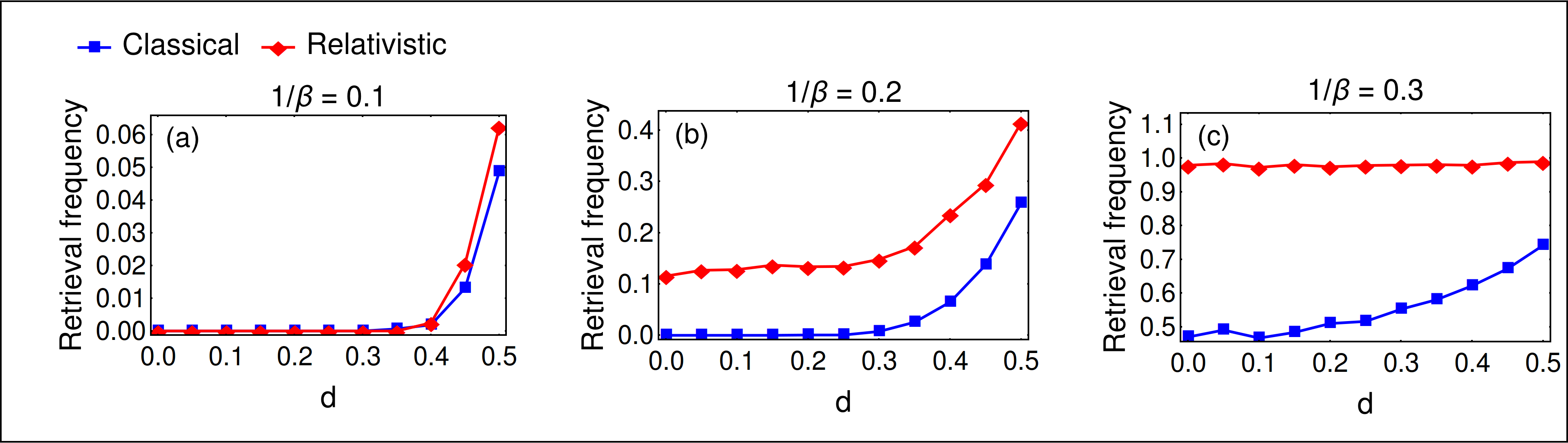}	
	\end{center}
	\caption{{\bfseries Stability of retrieval performances.} Results for retrieval frequency as a function of the initial spin-flip fraction $d$ for $\beta^{-1}=0.1$ (a), $0.2$ (b) and $0.3$ (c) in the range $0\le d \le 0.5$ (higher spin-flips fraction would take the initial condition immediately too far away from the spurious attractor, so they are not relevant in this discussion). In the left panel we see that for too low noise levels we have to significantly flip the neural states before both the models become capable to escape from the spurious state, however in the middle panel we see that already for mild noise level the relativistic model has the ability to drift away from the spurious toward the pure attractor, finally, increasing further the noise level in the network, in the right panel we can appreciate out the relativistic model out-performs w.r.t. its classical counterpart in the entire analyzed range.}\label{gardner}
\end{figure}

\end{itemize}

\subsection{Depth of the attractors and energy gaps}

Inspired by the pioneering works of the Gardner on the estimates about depth and stability of the basins of attractions of pure and spurious states \cite{Gardner1,Gardner1}, we now introduce the percentage of random spin-flips $d$ we impose on the system (such that for {\it e.g.} $d=0$ no random spin-flip at all is performed, while for {\it e.g.} $d=50\%$ we flip randomly one half of the total spins): the underlying idea is to start the system into a known state  ({\it e.g.} a spurious state) and then we reshuffle it by {\em kicking} randomly a percentage $d$ of its neurons and checking, as $d$ grows, the return (or the escape) of the network from the initial attractor. Results are shown in Fig. \ref{gardner} focusing on a 3-mixture spurious state in order to quantify the pruning capabilities of the relativistic model and compare them with those of its classical limit: while, for low noise level ($\beta^{-1} = 0.1$, left panel), we have to reach consistent percentages of spin-flip to allow the network to escape the spurious state ($d \sim 0.40)$ for both the models, already at mild noise level ($\beta^{-1} = 0.2$, middle panel) we see that spurious states already becomes instable in the relativistic extension -while being stable in the classical limit- and for values of $d \sim 0.35$ a further improvement in the pure state retrieval is obtained (for both the models). Finally in the last plot (right panel) we show that for moderate noise levels ($\beta^{-1} = 0.3)$ the relativistic model essentially always escape from the spurious state regardless of $d$, tacitely showing that the basins of attraction of those states are corrupted by the unlearning contributions nestled in the relativistic Hopfield cost function \eqref{relativistico}.
\newline
A reasonable cause of the increased performances of the relativistic Hopfield model may lie in a decreased energy barrier between the spurious states and the maxima surrounding them. To check this idea, again we can compare these gaps for the classical and the relativistic models: we still start the network's dynamics in a spurious state and end up in a pure state -the target pattern- but the way we move from to the spurious state to the target is via controlled noiseless random walks, {\it i.e.} a ground-state dynamics:  at each step, we select a spin $\sigma_i$ and if it is already aligned with the target pattern we want the network to reach we just move on, otherwise we flip it and we compute again the energy, then we repeat the process until the pure state is reached. Starting from a spurious state, and approaching a pure one, an energy barrier has to be crossed: we collected these energy gaps and, averaging  statistics consisting in $1600$ different runs (sampling $40$ different pattern choices to generate spurious states, for each of which we performed 40 different stochastic evolutions), and we found that the energy barrier to escape a spurious state is more than halved in the relativistic model if compared to the classical counterpart, {\it i.e.} $\Delta E_{relativ}/\Delta E_{classic} \sim 0.75$, hence confirming that unlearning has an effective role in pruning the network.

\section{A 1-parametric generalization of relativistic Hopfield model}
\begin{figure}[tb!]
	\begin{center}
		\centering
		\includegraphics[width=0.8\textwidth]{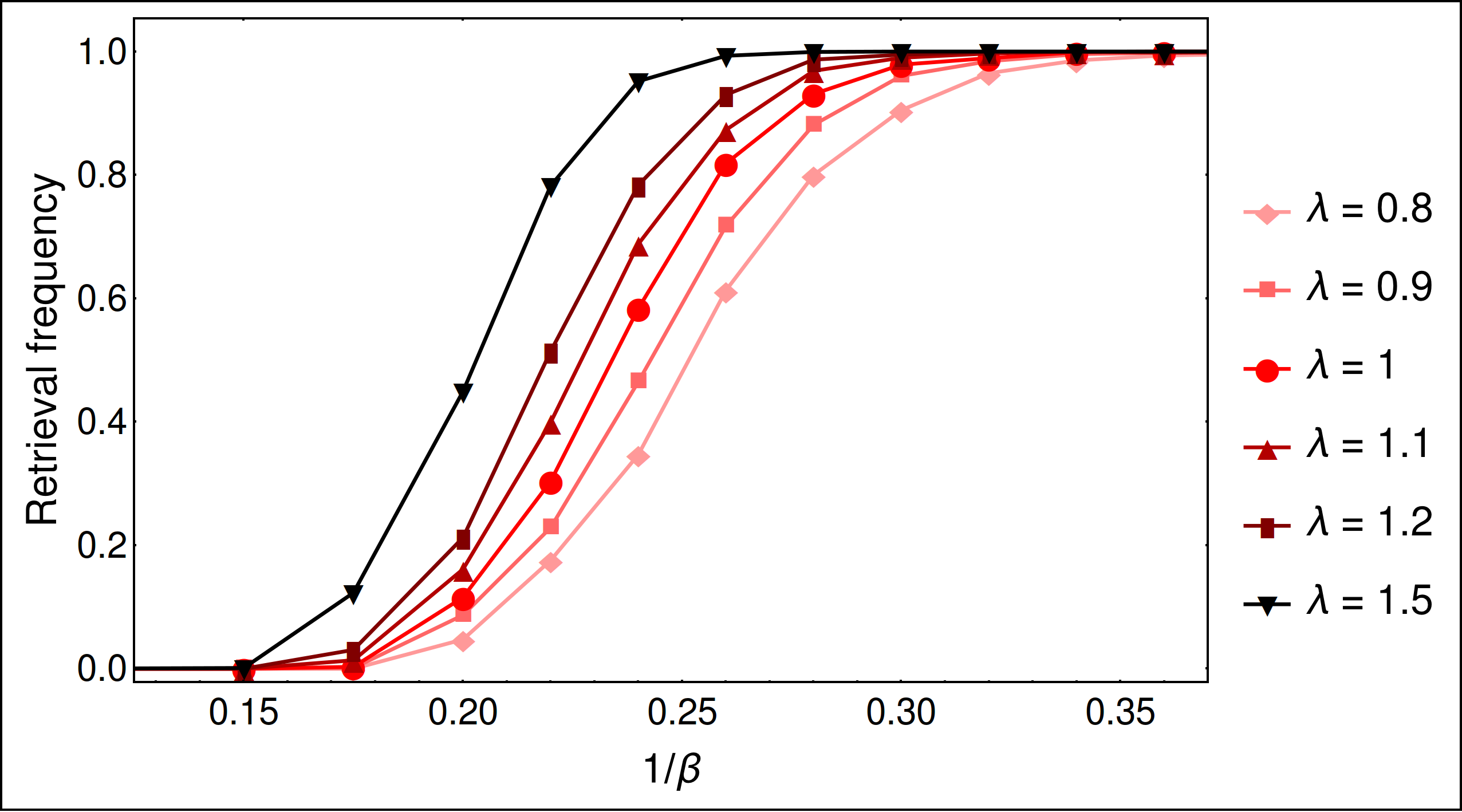}	
	\end{center}
	\caption{{\bfseries Retrieval performances for the 1-parametric model.} Results for 1-parametric model for a network of size $N=1000$ and $P=3$ patterns stored. The initial condition is aligned to the spurious state: $\sigma_{0,i}=\text{sign} (\xi^1_i+\xi^2_i+\xi^3_i)$. For values of the control parameter $\lambda$ lower than 1, performances gradually worsen with respect to the relativistic Hopfield model. On the other side, increasing the value of $\lambda$ beyond 1, the retrieval frequency accordingly grows and the network dynamics more likely ends in a pure state. As always for spurious state initial conditions, we averaged the results of the Glauber dynamics over 40 different pattern realizations and, for each of them, over 40 different stochastic evolution.}\label{lambdas}
\end{figure}

We would like to conclude this work with a possible continuation of our investigation. The relativistic extension of the Hopfield model has a nice justification in terms of the mechanical analogy through Hamilton-Jacobi formalism. On the side of statistical mechanics, our model is only one of the possible choices of a cost-function incorporating the principles of deep learning and network pruning, so it is reasonable to extend it with more general cost-functions which recover the Hopfield associative memory framework as the leading contribution in their Taylor expansion. A straightforward generalization of our proposal is indeed the 1-parametric Hamiltonian
\be
H_N(\boldsymbol{\sigma}\vert\boldsymbol{\xi})= -N \frac{\sqrt{1 + \lambda \bold m_{\ti N} (\boldsymbol{\sigma})^2}}{\lambda},
\ee
whose Taylor expansion
\be
\frac{H_{\ti N}}{N}=-\frac{1}{\lambda }-\frac{\bold m_{\ti N}^2}{2}+\frac{\lambda  \bold m_{\ti N}^4}{8}+O\left(\bold m_{\ti N}^{6}\right),
\ee
gives the Hopfield model at the leading order.\footnote{Of course, the zero-point energy $\lambda^{-1}$ does not affect the thermodynamic properties of the system.}
In order to have pruning corrections with alternate signes), we have to choose $\lambda$ to be a positive real number. To give a sketch of the performances of this new model, we compared the retrieval frequency curves for different $\lambda$ values for a network with $N=1000$ and $P=3$ and choosing the initial condition to be aligned to the spurious state, see Fig. \ref{lambdas}. As a result, for $\lambda$ higher than 1 (corresponding to the relativistic Hopfield model) lead to models performing better in retrieving when the network is prepared in a spurious state configuration. This is a good point which should encourage future study in this context. In particular, a more detailed investigation of this 1-parametric model (as well as more general suitable choices) is required, both numerically and analitically. We leave this point and the study of the high storage regime ({\it i.e.} replica trick analysis and the realization of the phase diagram) open for future works.

\section{Conclusions and further developments}

Regarding the unlearning phenomenon, quoting Hassibi and Stork \cite{pruning1}, {\em a central problem in machine learning and pattern recognition is to minimize the system complexity consistent with the training data. (...) If we begin with a trained network having too many weights, the question become: which weights should be eliminated?} Answering this question gave rise to {\em pruning} in machine learning. However, while pruning algorithms have been extensively exploited in neural networks away from detailed balance, {\it e.g.} radial-basis-function and/or feed-forward neural networks \cite{pruning2,pruning3,pruning4}, not so much has been said for stochastic neural dynamics obeying detailed balance, namely for associative neural networks as Boltzmann machines and/or Hopfield-like models. For these models, however, very powerful tricksappeared in the past to remove {\em spurious states} (locally stable states, pure pattern's mixtures that do not match exactly any of the stored patterns): an un-learning procedure inspired by REM-sleep has been proposed \cite{REM,Crick,HopfieldUnlearning} but it has always been seen as an {\em a fortiori} algorithmic procedure rather than an intrinsic property of a model.
\newline
Focusing on Deep Learning instead, since the seminal review by LeCun, Bengio and Hinton \cite{DL1}, where deep learning impressive skills have been listed alongside our limitation in understanding how deep learning machines achieve these scores, a plethora of contributions quickly appeared on the theme, among which the Hopfield's idea to use higher-order interactions (many-body terms) in the cost-function to be optimized. However, as for the unlearning, this intriguing proposal did not come out as a natural property of a models.
\newline
In this paper we tried to merge these two major breakthroughs in Artificial Intelligence   ({\it i.e.} pruning/unlearning and deep learning) by re-obtaining them as particular features of a unique and very natural model, the relativistic extension of the Hopfield paradigm.
\newline
To accomplish this task, at first, we had to develop a full mechanical analogy of the statistical mechanical treatment of the standard Hopfield model \cite{Hopfield},  checking that the mechanical analogy (largely in use in spin-glass Literature) does work correctly even when dealing with neural networks: in this mirror, the free energy of the model plays as a mechanical action and it obeys a classic  ({\it i.e.} not relativistic) Hamilton-Jacobi equation in the space of parameters (where the two-body coupling -coding for two points correlations- tunes the time, while the one-body coupling -coding for one point correlations- covers the role of space). Extremizing the action by the least square action principle we re-obtain the correct expression for the self-consistency of the Mattis order parameter, as obtained trough the standard route by Amit-Gutfreund-Sompolinsky \cite{Amit} in the low storage case.
\newline
Once checked that the mechanical analogy correctly recovers all the details of the celebrated Hopfield picture, we noticed two pathologies in the classical mechanical treatment, namely an underlying Minkowskian metric tensor and a bounded velocity for the fictitious particle of the mechanical motion. Both these observations suggested to extend the mechanical analogy to a relativistic treatment: this naturally introduced a novel cost-function, the relativistic generalization of the original Hopfield proposal.
\newline
This new model has been studied in all details, confined to the low storage (namely for regimes where there is abundance of available neurons w.r.t. patterns to be stored) and an analytical expression for its free energy, as well as a prescription for the evolution of the Mattis order parameters, have been explicitly obtained and shown to be in total agreement with numerical simulations. We also considered a 1-parametric generalization of the relativistic model, giving a numerical sketch of how performances increase with a suitable choice of the control parameter. This result opens the possibility to the study of more general cost-function choices.
\newline
Remarkably such Hamiltonians, once Taylor expanded, are shown to include all the (even) higher order monomials (beyond the two-body interactions of the original Hopfield framework) and these monomials succeed one another with alternate signs, such that the leading term is the standard pairwise Hopfield model (ensuring retrieval), the next one is a four-neuron coupling with reverse sign (hence appealing for unlearning/pruning the network), while the third term is a six-neuron coupling with the correct sign (hence appealing for deep learning/increasing memory storage).
\newline
These two features are recovered as {\em emergent properties} in the relativistic extension of the Hopfield model and clearly deserve detailed inspection: in this paper we deepened solely the former (namely the unlearning capability that the network uses for removing spurious states from the retrieval landscape), postponing an analysis of the storage capacity of the network (the high storage case) to a forthcoming paper. The reason behind this choice is that the required mathematical treatment is completely different. When dealing with the high storage (i.e to check storage capabilities and deep learning skills trough disordered statistical mechanics)  further concepts borrowed from the statistical mechanics of spin glasses are required (mainly replicas and overlaps among replicas). We plan to report soon also results on the high-storage regime.
\newline
A conclusive remark is that while we restricted our analysis to binary weights/patterns (mainly for the sake of continuity with the mainstream in Literature on Hopfield networks), our approach holds for real-valued variable too (as can be easily understood noticing that the pattern's average is performed at the end of the whole calculation, en route for the explicit expression of the model's free energy).
\newline
Finally, we conclude this paper by observing also that the mechanical analogy can perform as a useful tool possibly beyond the associative memory theory, for other problems in Artificial Intelligence, Hopfield networks remaining its first benchmark.

\section*{Acknowledgement}
AB acknowledges partial financial support by GNFM-INdAM (via AGLIARI2016), by  MIUR (via research basic fundings) and by MIUR (via Calabria project, name to be added).
\newline
AF would like to thank Dr. Elena Agliari for the useful discussions, especially concerning the numerical analysis.

\end{document}